\documentclass[onecolumn,11pt,draftcls]{IEEEtran}
\usepackage{amsmath} % assumes amsmath package installed
\usepackage{amssymb}  % assumes amsmath package installed
\usepackage{psfrag,algorithm, algorithmic, epsfig, verbatim}

\newtheorem{definition}{Definition}[section]
\newtheorem{theorem}{Theorem}
\newtheorem{lemma}{Lemma}

\title{Intent Inference and Syntactic Tracking   with GMTI Measurements \thanks{A short version of this paper appears
in the Fusion 2009 conference in July 2009.}}

\author{Alex Wang,
Vikram Krishnamurthy~\IEEEmembership{Fellow,~IEEE},
Bhashyam Balaji
\thanks{Vikram Krishnamurthy and Alex Wang are with the Department of Electrical and Computer Engineering, Univ. of British Columbia, Vancouver, Canada, V6T 1Z4. email: vikramk@ece.ubc.ca and alexw@ece.ubc.ca. B. Balaji is with Defence R\&D  Canada Ottawa, ON, Canada. email: bhashyam.balaji@drdc-rddc.gc.ca}}

\begin{document}

\maketitle

%\tableofcontents

\begin{abstract} In conventional target tracking systems, human operators use the estimated target tracks
 to make higher level inference  of the target behaviour/intent.
 This paper develops syntactic filtering algorithms that assist human operators by extracting  spatial patterns from target tracks to identify  suspicious/anomalous spatial trajectories. The targets' spatial trajectories are modeled by a stochastic context free grammar (SCFG) and a switched mode state space model. Bayesian  filtering algorithms for stochastic context free grammars are presented for  extracting the syntactic structure and illustrated for a
 ground moving target indicator (GMTI) radar example. The performance of the algorithms is tested with the experimental data collected using DRDC Ottawa's X-band Wideband Experimental Airborne Radar (XWEAR).
 \end{abstract}

\begin{IEEEkeywords}
Ground Moving Target Indicator (GMTI), 
Stochastic Context Free Grammar (SCFG),
Space-Time Adaptive Processing (STAP), Stochastic Parsing, Intent tracking, Bayesian Inference
\end{IEEEkeywords}

\section{Introduction} 

\subsection*{Context and Main Results}

For tracking ground-based maneuvering targets, conventional tracking systems deal with the following switched mode state space model \cite{BL93,LJ05,BP99}
\begin{align}
x_{k} = & F (a_k) x_{k-1} +  v_{k-1}(a_k)  \nonumber \\
z_k = & h(x_k)+ w_k . 
\label{GMTI:eqn:statespace}
\end{align}
Here $k$ denotes discrete time, $x_k$ denotes the kinematic target state such as position and velocity, and $z_k$ denotes the sensor detections (observations). The random processes $v_k$ and $w_k$ denote the state and observation noise respectively. The mode sequence $a_{1:k}= \{a_1,\ldots,a_k\}$ summarizes a sequence of maneuvers or modes that causes the ground-based target to move in a two dimensional spatial trajectory. Conventional tracking of maneuvering targets assumes that the mode sequence $a_{1:k}$ is a finite state Markov chain, and aims to compute the posterior distribution $P(x_k, a_k|z_{1:k})$ so as to compute conditional mean estimates of $x_k$ and $a_k$. This is typically done by a state-of-the-art tracking algorithm involving particle filters, Interacting Multiple Models (IMM), and variable structure IMM (VS-IMM) \cite{BL93,RAG04,KBP00}. (In VS-IMM, the kinematic model of the moving objects depend on the road direction and the terrain type). These Bayesian recursions exploit the Markovian assumption of the mode sequence $a_{1:k}$ to estimate $x_k,a_k$.

Motivated by intent-inference
%{\em GMTI} (ground moving target indicator) 
applications, this paper  deals with a higher level of abstraction which we call \textit{Syntactic Tracking}. Suppose we are interested in whether a target is circling a restricted area (perimeter surveillance), or alternatively if a vessel is loitering near the coast (for a possible smuggling attempt). In such cases, the human operator is primarily interested in determining specific patterns in target trajectories from estimated tracks. These patterns can then be used to infer the possible intent of the target \cite{BP99}.
%Moreover, in military setting, a forward quarter intercept or a pincer is characterized by two cooperating vehicles maneuvering in arcs \cite{Blackman1999}. 
%However, in such cases, the operator is not only interested in kinematic estimates of position and velocity, he is also determining specific patterns in target trajectories from conventional track estimates, which he can use as aid to infer the possible intents of the targets.
Examples of such specific patterns include loops, arcs, circles, rectangles, and combination of these, and they exhibit complex spatial dependencies. 
%As is well known in syntactic pattern analysis \cite{Fu82}, the Markovian assumption of $a_{1:k}$ in (\ref{GMTI:eqn:statespace}) is incapable of modeling such complex spatial patterns.
 The key modeling contribution of this paper is to construct a syntactic model to characterize various spatial patterns with a linguistic construct called \textit{stochastic context free grammar (SCFG)}.
%using terminology from formal language, we will demonstrate that $a_{1:k}$ in (\ref{GMTI:eqn:statespace}) is not a regular language (equivalent counterpart to Markovian state sequence), but a context free language. Context free language can be compactly characterized by a stochastic context free grammar (SCFG), which is 
% T
Thus the main goal is to devise SCFG models and associated polynomial time Bayesian syntactic parsing algorithms to extract spatial patterns from the mode sequence $a_{1:k}$ estimated by the conventional target tracker. In other words, \textit{this paper develops models and automated syntactic filtering algorithms to assist the human operator in determining specific target patterns}.
The algorithms presented in this paper use the track estimates from an existing  tracker to perform syntactic
filtering. In this sense, they are at a higher layer of abstraction  than conventional tracking and are  fully compatible with existing trackers, see Fig.\ref{GMTI:fig:framework} for a more detailed
schematic. Indeed, it is not the intent
of this paper to re-design conventional target tracking which is a well trodden area.

\subsection*{Why Use Stochastic Context Free Grammars (SCFGs)?}

In formal language theory,
grammars can be classified into four different types depending on the forms of their production rules \cite{DEKM98}. Stochastic regular grammars or finite state automata are equivalent to 
HMMs. SCFGs (which will be defined in Sec.\ref{subsec:scfgbackground}) are a significant generalization of regular grammars. Only stochastic regular and SCFGs have polynomial complexity estimation algorithms and are therefore of practical use in radar tracking applications. It is well known in formal language theory, that SCFGs are more general than HMMs (stochastic finite automata) and can capture long range dependencies and recursively embedded structures in patterns.

%\subsubsection*{Implementation Advantages}
The implementation of the syntactic filtering system with SCFG has several potential advantages:\\
\textup{(i). {\em User-friendly Models}}: SCFG have a compact formal representation in terms of production rules that can permit human radar operators to easily codify high-level rules,
see \cite{VKWH07,WK08} where the complex dynamics of a multifunction radar were
modeled using SCFGs. In this paper, it allows us (and radar engineers)
 to model complex spatial patterns of target trajectories such
 as if a target is circling a building or intersecting in trajectory with another target.
This then permits the design of high-level Bayesian signal processing algorithms
to estimate such trajectories.   The ability for the designer to encode knowledge is important because the lack of field data in a defence setting often hinders the application of Bayesian filters as they  require
substantial amounts of  training data. \\
(ii) {\em Ability to Model Complex Spatial Trajectories}: The recursive embedding structure of the possible target geometric patterns is more naturally modeled in SCFG. As will be shown later, the Markovian type model has dependency that has variable length, and the growing state space is difficult to handle since the maximum range dependency must be considered. \\
(iii) {\em Predictive Power}: SCFGs are more efficient in modeling hidden branching processes when compared to stochastic regular grammars or hidden Markov models with the same number of parameters. The predictive power of a SCFG measured in terms of entropy is greater than that of the stochastic regular grammar \cite{LY90}. SCFG is equivalent to a multi-type Galton-Watson branching process with finite number of rewrite rules, and its entropy calculation is discussed in
\cite{MS92a}.

\subsubsection*{Main Results} 
For simplicity, our setting is for targets that move in two dimensional space,
and airborne {\em GMTI} (ground moving target indicator) radar is used as the primary sensing  platform throughout the paper. However, the syntactic filtering results of this paper can be
used with other sensor technologies such as multiple video/imaging sensors, etc.
Because of the vast amount of data  generated by GMTI trackers, there is strong motivation to develop automated algorithms that yield a high level interpretation from the tracks. The main results of the paper are:

\noindent
\textit{1. Combined Tracking and Trajectory Inference:}
Sec.\ref{sec:overview} sets the stage by describing our entire framework for 
syntactic filtering using  conventional track estimates. We review SCFGs, formulate the elementary modes that lead to trajectories such as arcs and modified rectangles, and describe how  syntactic
tracking fits into a complete tracking system.

\noindent
\textit{2. SCFG Modulated State Space Model:}
Sec.\ref{sec:GMTImodeling}  presents a SCFG modulated state space model that permits modeling of complex spatial trajectories. We derive probabilistic production rules that characterize the target motion patterns, and present a detailed structural analysis of the SCFG model. Using
 formal language techniques and the Pumping Lemma \cite{HMU07}, we show specific syntactic pattern like an arc generates a context free language, and it cannot be modeled by Markov models efficiently. Moreover, the well-posedness of the syntactic model is studied based on the branching rate of the model, and conditions over which the language distribution is proper are given, i.e. the  conditions that ensure the distribution of the language generated by the model sums to one.

% add meaning of well-posedness

\noindent
\textit{3. Bayesian Syntactic Filtering:}
Sec.\ref{sec:GMTIparsing} presents the Bayesian syntactic filtering algorithm. The interpretation of the syntactic patterns are represented by parse trees built on top of the target trajectories, which is tracked at the detection level by Bayesian filters such as particle filter and IMM/extended Kalman filter \cite{KBP00}, and at the mode level by a generalized Earley Stolcke Bayesian parser 
\cite{LBK04}. The Earley Stolcke algorithm is a generalization of the Forward-Backward algorithm for Hidden Markov Models (HMM), and it allows real time forward parsing. The complexity of the algorithm is $O(l^3)$, where $l$ is the length of the input string. 
%The tracking algorithms are based mainly on the state-space model, where the kinematics of a single or multiple targets are recursively estimated with algorithms such as Kalman, HMM, or VS-IMM \cite{KBP00,LBK04}; VS-IMM is the more sophisticated model where the kinematic model of the moving objects depend on the road direction and the terrain type. 

% add complexity and performance

\noindent
\textit{4. Experimental Validation of Syntactic Filtering:}
Sec. \ref{sec:simulation} gives a detailed experimental analysis of the syntactic filtering algorithm on a real life GMTI example. The GMTI data was collected using the DRDC Ottawa's X-band Wideband Experimental Airborne Radar (XWEAR)\cite{DMH03,BD06}, and numerical studies of the syntactic filtering algorithms are performed using the data. The experimental results show that syntactic tracker not only accurately estimates the target's trajectory pattern, but also can be used to improve the accuracy of conventional trackers.

\subsection*{Literature Review}

SCFGs have widely been used in language processing. The complexity of the language in sentence structure and grammatical dependency made state space models such as linear predictive coding \cite{Col05} and hidden Markov model \cite{Jel97} inadequate, and the application of  stochastic grammar in language modeling has been researched extensively, where its syntax naturally models the language's grammar structure \cite{MS99a}. In addition to language processing, SCFG has been a major computational tool in biology for DNA and RNA sequencing \cite{DEKM98}. Because of the three-dimensional folding of the proteins and nucleic acids, HMM becomes insufficient, and SCFG is essential for capturing
the long range dependencies of  spatial folding.

\textbf{SCFG in Tracking}: In conventional tracking, effort has been spent to enhance the tracker by incorporating information other than the kinematic states. In \cite{BP99}, attribute tracking is discussed where target class information such as wing span and jet engine modulation are utilized for data association. In \cite{GD06}, features in targets' path trajectory, velocity, and radar cross section are used for target and track classification. In contrast to attribute tracking and target track classification, the syntactic models not only can deal with static features, but they are also particularly suitable to finding patterns in mode sequences with complex multi-scale structure and recursive nature. For example, in plan recognition,  plans of an agent, typically the actions, have to be inferred from observations. \cite{Cha93} approached the problem with Bayesian network, but due to the complex structure generating the actions, it is too computationally intensive.  In addition, in video surveillance, hierarchical hidden Markov model is applied to track sequences of human actions \cite{LBVW03}, and it can be shown that the hierarchical hidden Markov model is a special case of SCFG \cite{FST98}. SCFG can be applied directly to establish high level inferences from primitives generated from observations. In \cite{IB00}, SCFG is applied to detect sequences such as dropping a person off or picking a person up in a parking lot. Moreover, in \cite{LOSA06}, movements of targets such as U-turns are inferred based on measurements collected from a sensor network. For those SCFG based tracking, the focus is on the high level inference, and the coupling between the high level inference and the Bayesian tracking is typically very loose, i.e. $a_{1:k}$, are independently generated from sensor measurements, and the temporal constraints are imposed only at the higher inference level. 

%The aim of the syntactic filtering system is to form tracks from multiple frames of GMTI detections, and infer geometric patterns from the track estimates. Under this methodology, the syntactic filtering problem is reduced to a pattern recognition problem, which, in general, can be classified into two major categories: decision theoretic and syntactic pattern recognition \cite{Fu82}. The decision theoretic approach extracts features from patterns, and patterns are classified according to a partitioned feature space. The syntactic approach, on the other hand, decomposes a pattern to its constituent sub-patterns, and recognizes the pattern by building its syntactic structure. The syntactic approach that describes a complex pattern with simpler sub-patterns in a hierarchical fashion is very attractive, because the knowledge of sub-patterns is often difficult to learn with data, but is priorly known by domain experts. However, because of the problem complexity, no single approach is sufficient. At the meta level, because of the temporal dependency in the trajectory measurement, and lack of training data, pure decision theoretic approach is not practical as the number of features required is often very large. At the sensor detection level, the frequency and the amount of the GMTI detections make syntactic approach unfeasible due to the computational intensity. The system framework of the syntactic filtering consists of both the decision theoretic and the syntactic approaches, and it is discussed in this section.

\textbf{GMTI}: Conventional single-channel radars deployed to perform ground surveillance are limited in the sense that they are only capable of performing detection of fast movers, and identification of stationary targets via SAR imaging algorithms. GMTI radar with space-time adaptive processing (STAP) enables the near-real time detection of ground moving objects over a large area. STAP is a generalization of adaptive array signal processing techniques based on the Wiener filter \cite{Kle98}, and it incorporates techniques such as eigenvector projection and the least-squares method. In conventional adaptive array signal processing, a Wiener filter is formed for a signal vector whose components are the signals received at multiple apertures from a single pulse. In STAP, on the other hand, the Wiener filter is formed for a received signal vector whose components are some function of signals received at multiple apertures, which are moving, for more than one pulse. In other words, STAP provides a two-dimensional adaptive filter where the apertures and pulses furnish the spatial and temporal samples. It is noted that although STAP-based GMTI is considered here, the techniques developed can be used in conjunction with other detection techniques, such as detection algorithms in the image domain, i.e., synthetic aperture radar (SAR) based GMTI algorithms.

\section{Overview of GMTI Based Syntactic Tracking}
\label{sec:overview}

To motivate the syntactic  modelling and syntactic tracking algorithms presented in this paper,
in this section we present an overview of our  approach to syntactic tracking.
 {\em Our premise for syntactic tracking is that the geometric pattern of a target's trajectory can be modeled as "words" (mode sequence) spoken according to a SCFG language}. So the  intent or behaviour of the targets can be determined by SCFG signal processing methods (syntactic pattern recognition techniques). The basic idea of the syntactic pattern recognition is that complex patterns can be expressed as simpler patterns.  That is, we decompose high level descriptors of target intents into motion trajectories consisting of a fixed set of primitive geometric patterns such as a line or an arc, and the primitive geometric patterns into kinematic modes that can be estimated by a target tracker. In this section, some examples of syntactic tracking are discussed, and the system framework that supports syntactic tracking is presented.
 
 \subsection{Examples}
 
In this paper, we illustrate the syntactic tracking algorithms
with examples from GMTI radar.
 Based on these GMTI detections, the aim is to construct an algorithm for continuous ground surveillance that infers the meta description of the moving units by classifying and labelling their trajectories according to their geometric patterns. Consider the following examples that motivate our approach to syntactic tracking.

{\em 1. Syntactic tracking in threat inference}: A vehicle approaches a security gate of a building and turns around. It then circles around the perimeter of the building in the midst of other moving vehicles. Given GMTI track information of multiple moving vehicles, how can this behaviour be recognized as a threat? Equivalently, how can a threat be associated with the complex spatial trajectory of Òmaking a U-turn and then circling a buildingÓ, and how can the spatial trajectory be identified from geometric patterns?

\begin{figure}
\centering
\includegraphics[width=0.6\linewidth]{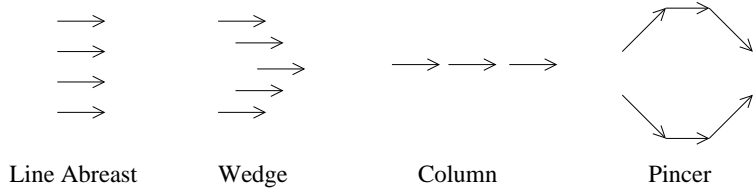}
\caption[The battalion formations]{The battalion formations. Line abreast and wedge are offensive 
combat formation, column is a traveling technique, and pincer is a
intercepting technique.}
\label{GMTI:fig:formation}
\end{figure}

{\em 2. Syntactic tracking in military operations}: Fig. \ref{GMTI:fig:formation} illustrates examples of high level descriptions of motion patterns that are common in military ground surveillance, where each is characterized by certain combination of geometric patterns \cite{Car01}; the line abreast and wedge formation are offensive combat formations with each vehicle moving in linear trajectory; pincer, on the other hand, consists of two vehicles maneuvering in mirroring arc trajectories. With this high level description, inferences can be made to determine if the ground units are in offensive, defensive or reconnaissance operation.

\subsection{Syntactic Target Tracking System Framework}
\label{subsec:architecture}

Let $\mathcal{M}$ denotes the set of geometric patterns of interest. For simplicity, we consider
 \begin{equation}
\mathcal{M} = \{ \text{line}, \text{arc}  , \text{m-rectangle}\},
\label{eqn:m}
\end{equation}
and these geometric patterns are described later in detail in Sec.\ref{subsec:scfgandtracking}.
Syntactic filtering is built on top of multiple model approach to target tracking, and it enables the characterization and identification of geometric patterns from the target trajectory. The main stream multiple model approach is the interacting multiple model (IMM) \cite{BLK01}, and it recursively computes the state information with the following distribution function 
\begin{equation}
P(x_k|z_{1:k}) = \sum_{a_k} P(x_k | a_k,  z_{1:k})P(a_k | z_{1:k}).
\label{eqn:state}
\end{equation}
In IMM formulation, the exponentially growing number of mode sequences is approximated by merging the $r^2$ hypotheses at each instance to $r$ hypotheses, where $r$ is the number of modes \cite{LJ05}. However, because of the merging, the geometric information that could be used for higher level intent inference is lost. Instead of merging, syntactic filtering keeps the mode sequence, and applies pruning to keep the computation manageable.

More specifically, the syntactic filtering is only applied to the  second term in (\ref{eqn:state}), the mode probability. In order to estimate its value, only the most likely mode sequence is kept, and, using Bayesian model averaging, the probability is computed approximately as
\begin{align}
P(a_k|z_{1:k}) = & \sum_{l \in \{RG, CFG\}} \sum_{a_{1:k-1}} P(a_k, a_{1:k-1}, G^l | z_{1:k}) \nonumber \\
\approx & P(a_k, a^*_{1:k-1}|G^{CFG}, z_{1:k})P(G^{CFG} | z_{1:k}) + \sum_{a_{k-1}} P(a_k, a_{k-1}|G^{RG},z_{1:k})P(G^{RG}|z_{1:k})
\label{eqn:feedback}
\end{align}
where $a^*_{1:k-1}$ is the most likely mode sequence given the SCFG model (as $a_{1:k} \in \mathcal{L}_{CFG}$ models geometric patterns of the target trajectory), and the second term is the conventional IMM tracker. 
Given the track estimates, syntactic filtering allows classification of the mode sequence into geometric patterns. The maximum a posterior (MAP) pattern is then computed as
\begin{equation}
\hat{m} = \arg \max_{m \in \mathcal{M}} P(a^*_{1:k}|G_{m}),
\label{GMTI:eqn:meta}
\end{equation}
where  $G_m \in G^{CFG}$ is the SCFG of the geometric pattern $m \in \mathcal{M}$. The computation of the associated probabilities is discussed in Sec. \ref{sec:GMTIparsing} where the SCFG parsing algorithm that performs the syntactic analysis is described.

\begin{figure}
\centering
\psfrag{GMTIout}{$z_k$}
\psfrag{Trackerout}{$P(a_k|z_{1:k}, G^{RG})$}
\psfrag{Parserout}{$\hat{m}$}
\psfrag{Kinematicout}{$\{\hat{x}_k, \hat{a}_k\}$}
\psfrag{Parserfeedback}{$P(a_k, a^*_{1:k-1}|G^{CFG})$}
\includegraphics[width=0.9\linewidth]{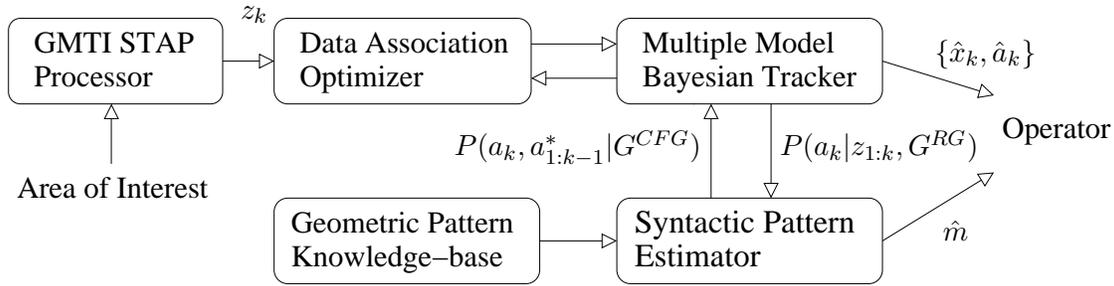}
\caption{The system framework for the GMTI based syntactic filtering system. The GMTI sensor measurements are denoted by $z_k$, the kinematic states by $x_k$, and the modes by $a_k$. $G^{RG}$ refers to the Markov model (regular grammar) characterizing the mode transitions, and $G^{CFG}$ refers to the context free model characterizing the geometric patterns.}
\label{GMTI:fig:framework}
\end{figure}

Given this formulation, the system framework of this syntactic filtering system is summarized in Fig. \ref{GMTI:fig:framework}. The system framework consists of five components, and their functionalities are described as follows: The GMTI STAP processor detects ground moving targets and returns their estimated range, angle, and range rate. The data association optimizer assigns sensor measurements to tracks. The multiple model Bayesian tracker keeps track of the detected targets, and recursively computes the targets' kinematic states and their mode probabilities given the sensor measurements. The geometric pattern knowledge-base stores the prior knowledge of the relevant motions in terms of production rules. Build on top of the conventional multiple model Bayesian tracker, the syntactic pattern estimator (stochastic parser) infers geometric patterns from vehicle's trajectory, and provides feedback to track estimate in terms of mode probability estimation to enhance tracking accuracy.

\vspace{5pt}
\noindent
{\em Remark}: Various techniques already exist to perform data association. The joint probabilistic data association (JPDA) algorithm that evaluates the measurement-to-track association probabilities \cite{LBK04}, the multiple hypothesis tracking (MHT) algorithm that enumerates all feasible measurement-to-track hypotheses \cite{BP99}, and the assignment algorithms  that solve data association as a constrained optimization problem are all relevant techniques in this field. The focus of the paper is on the syntactic interpretation of target trajectories, and because the assignment algorithms are more modular in the sense that they can work with different tracking algorithms, for example IMM and VS-IMM, they are well suited to deal with the data association problem in this paper. \cite{LBK04} not only solves the data association problem, but also the tracking of move-stop-move targets.

%Sec. \ref{subsec:dataassociation} will discuss the assignment based data association.

\section{Syntactic Modeling for Ground Surveillance}
\label{sec:GMTImodeling}
Given the overview of our approach presented above,
this section presents complete details on the syntactic modelling of target trajectories using SCFGs.
The background on SCFG is provided in Sec. \ref{subsec:scfgbackground}. Sec. \ref{subsec:GMTIprimitive} discusses the state space models that estimate the mode sequence from GMTI detections, Sec. \ref{subsec:scfgandtracking} and \ref{subsec:SCFGformulation} present the syntactic modeling of the geometric patterns with SCFG, and finally, Sec. \ref{subsec:SCFGanalysis} proves  the well-posedness of the SCFG model (in terms of ability to model specific patterns). This section
thus sets the stage for Bayesian algorithms (parsing algorithms) to classify the target trajectory
and hence the target's intent that are presented in Sec.\ref{sec:GMTIparsing}.

\subsection{SCFG Background}
\label{subsec:scfgbackground}

With the motivation outlined above, we will use SCFGs to model  geometric spatial patterns of target trajectories. Since SCFGs are not widely used in radar signal processing,  we begin with a short formal description of  SCFGs and a summary of syntactic analysis (syntactic parsing). In formal language theory, a grammar $G$ is a four-tuple $<\mathcal{N},\mathcal{T}, \mathcal{P}, S>$ \cite{DEKM98}. Here $\mathcal{N}$ is a finite set of nonterminals, $\mathcal{T}$ is a finite set of terminals, and $\mathcal{N} \cap \mathcal{T} =\emptyset$. $\mathcal{P}$ is a finite set of probabilistic production rules, and $S\in \mathcal{N}$ is the starting symbol. As will be shown later in generation of a parse tree, nonterminals are the nodes that may generate other nonterminals and terminals, and terminals are the leaves. Throughout the paper, lower case letters are used to denote terminals, and upper case letters nonterminals. Greek letters are used to denote concatenated strings of terminals and nonterminals.

\begin{definition} \label{def:srg}
\textbf{[Stochastic Regular Grammar]} Stochastic regular grammars, denoted as $G_{RG}$, are equivalent to hidden Markov models (with termination state $\in \mathcal{N}$) and have production rules of the form $A \rightarrow a A$ and $A \rightarrow a$ with probabilities $P(A \rightarrow aA)$ and $P(A \rightarrow a)$ specified, where $A \in \mathcal{N}$. $\mathcal{N}$ corresponds to the state space of the hidden Markov model, and $\mathcal{T}$ corresponds to its observation space. The set of all terminal strings generated by regular grammar is called the regular language and it is denoted as $\mathcal{L}_{RG}$.
\end{definition}

\begin{definition} \label{def:scfg}
\textbf{[Stochastic Context Free Grammar]} SCFG, denoted as $G_{CFG}$, have production rules, $\mathcal{P}$, of the form $A \rightarrow \eta$ with probabilities $P(A \rightarrow \eta)$ specified, where $A \in \mathcal{N}$ and $\eta \in (\mathcal{N} \cup \mathcal{T})^+$. $(\mathcal{N} \cup \mathcal{T})^+$ denotes the set of all finite length strings of symbols in $(\mathcal{N} \cup \mathcal{T})$, excluding strings of length 0 (the case where length 0 string is included is indicated by $(\mathcal{N} \cup \mathcal{T})^*$). The set of all terminal strings generated by SCFG is called context free language and it is denoted as $\mathcal{L}_{CFG}$. The grammar is context free because the left hand side of its production rule only has a single nonterminal (independent of its context). To contrast, a grammar is context sensitive if it has production rules of the form $\rho_1 A \rho_2 \rightarrow \rho_1 \eta \rho_2$, where $\rho_1, \rho_2 \in (\mathcal{N} \cup \mathcal{T})^*$ and $\eta$ cannot be empty.

A context-free grammar is self-embedding if there exists a nonterminal $A$ such that $A \stackrel{*}{\Rightarrow} \eta A \beta$ with $\eta,\beta \in (N \cup T)^+$. A self-embedding SCFG cannot be represented by a Markov chain \cite{Fu82}. 
\end{definition}

\begin{figure}
\centering
\includegraphics[width=0.7\linewidth]{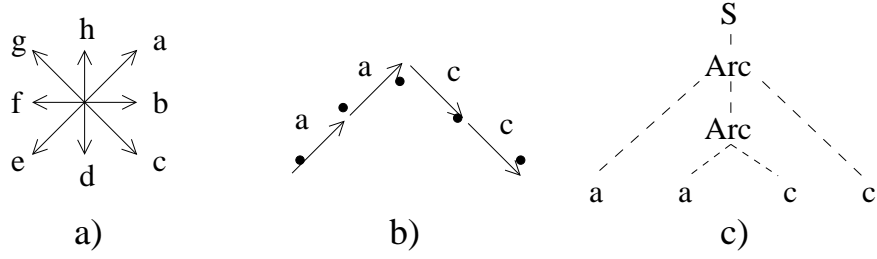}
\caption[Syntactic analysis of estimated tracks]{a) Building blocks of the trajectory. b) A sample trajectory and the estimated modes. c) Syntactic analysis of the sequence of estimated modes.}
\label{GMTI:fig:track1}
\end{figure}

\vspace{3pt}
\noindent
{\em SCFG Example}: Let the set of terminals be $\mathcal{T}=\{a,b,\ldots,h\}$ as illustrated in Fig. \ref{GMTI:fig:track1}a), and they represent the direction of travel of a target. A target trajectory is shown in Fig. \ref{GMTI:fig:track1}b), and it can be compactly expressed as a string of terminals $aacc$. Fig. \ref{GMTI:fig:track1} c) demonstrates one likely generation of terminals from the hypothesis that the pattern is an arc, and how segments of the string is ``explained" by nonterminals that comprise it. The set of nonterminals in this example are $\mathcal{N} = \{ \text{Arc} \}$, and the production rules used are 
\begin{align*}
\text{Arc} \rightarrow & \text{a ~Arc~ c $|$ a~ c }
\end{align*}
The symbol $\rightarrow$ indicates ``replace with", and the symbol $|$ indicates ``or". Suppose we have a concatenated string $xA$,  where $x$ is any combination of nonterminals and terminals, and $A$ is a nonterminal, a one step derivation using the rule $A \rightarrow a A$ yields $xA \rightarrow xaA$. The derivation process of  the example in Fig. \ref{GMTI:fig:track1} can be expressed as a iterative application of the production rules, as shown below:
\begin{align*}
\text{S} & \rightarrow \text{Arc}  \rightarrow \text{a Arc c}  \rightarrow \text{a a c c}
\end{align*}

\subsection{State Space Model for Target Trajectory}
\label{subsec:GMTIprimitive}

\begin{comment}
Below we formulate the kinematic models for the multiple model Bayesian tracker that can capture the modes that are illustrated in Fig. \ref{GMTI:fig:track1} a). The formulation of these modes are based on constant velocity model with directional process noise \cite{RAG04,KBP00}. However, it should be noted that in order to support the higher level geometric pattern recognition, the number of modes is kept finite, and tradeoff has to be made between tracking accuracy (model mismatch) and parsing complexity (number of terminals, or modes, and nonterminals).

Standard kinematic models assume equal variance for the process noise in all unit directions to allow for the target to move with equal probabilities among the unit directions. To model the modes, in this paper the process noise is assumed to have different noise variance \textit{along} and \textit{perpendicular} to the direction of the modes. If we know the ground target is moving along a particular direction, then the covariance perpendicular to the direction should be small. 

Let $k$ denotes discrete time, the assumed target dynamics is  $$\mathbf{x}_k = F\mathbf{x}_{k-1} + Gv_{k-1}(a_k).$$ $\mathbf{x}_k = (x_k, y_k, \dot{x}_k, \dot{y}_k)' $ denotes the ground moving target's position and velocity in Cartesian coordinates, and assuming constant velocity model, the transition matrix model and the noise gain are, respectively,
\end{comment}

Let the set of terminals $\mathcal{T} =  \{a,b,c,d,e,f,g,h\}= \{\pi/4, \pi/2, 3\pi/4, \pi, 5\pi/4, 3\pi/2, 7\pi/4, 2\pi\}$ denote the 
possible directions of travel of the moving target. Fig.\ref{GMTI:fig:track1}a illustrates these 8 possible acceleration
directions of the target depicted by the terminals $a,b,\ldots,h$.

At each time $k$, $a_k \in \mathcal{T}$ denotes mode of the target.
The target dynamics are modelled as 
\begin{equation}
\mathbf{x}_k = F\mathbf{x}_{k-1} + Gv_{k-1}(a_k).
\label{eq:main}\end{equation}
 $\mathbf{x}_k = (x_k, y_k, \dot{x}_k, \dot{y}_k)' $ denotes the ground moving target's position and velocity in Cartesian coordinates, and assuming constant velocity model, the transition matrix model and the noise gain are, respectively,$$ F = \left(\begin{array}{cccccc}
1 & 0 & T & 0 \\
0 & 1 & 0 & T  \\
0 & 0 & 1 & 0 \\
0 & 0 & 0 & 1
\end{array} \right),
G = \left(\begin{array}{ccc}
T^2/2  & 0 \\
0 	   & T^2/2 \\
T 	   & 0 \\
0 	   & T \\
\end{array} \right).$$
The process noise $v_{k}$ is a white Gaussian process with the covariance matrix 
$$\mathcal{Q} =
\rho_{a_k}\cdot
\left(\begin{array}{cc}
\sigma_a^2 & 0 \\
0 & \sigma_o^2
\end{array}\right)
\cdot \rho_{a_k}^\prime,
\mbox{with } \rho_{a_k} =
\left(\begin{array}{cc}
\sin a_k & \cos a_k \\
-\cos a_k & \sin a_k
\end{array}\right),$$ where $^\prime$ denotes transpose, and
$\sigma_a^2$ is the uncertainty along the direction indicated by $a_k$ and $\sigma_o^2$ is orthogonal to it. Thus the  modes $a_k$ modulate the process noise $v$ and cause
it to switch between different variance values.

{\em Remark}: The above  model is more suitable for ground targets compared to acceleration models (e.g. mean adaptive acceleration models and the semi-Markov jump process models) since  ground moving vehicles do not exhibit such maneuverability. 
Standard kinematic models assume equal variance for the process noise in all unit directions to allow for the target to move with equal probabilities among the unit directions. To model the modes, in this paper the process noise is assumed to have different noise variance \textit{along} and \textit{perpendicular} to the direction of the modes. If we know the ground target is moving along a particular direction, then the covariance perpendicular to the direction should be small.

The observation model describing the output of the GMTI STAP measurements is 
\begin{align} 
z_k & = h(\mathbf{x}_k) + w_k \nonumber \\
h(\mathbf{x}_k)& =
\left[ \begin{array}{c} 
	r_k \\ 
	\dot{r}_k \\ 
	\theta_k \end{array} \right]=
\left[ \begin{array}{c} 
	\sqrt{{\bar{x}_k}^2+{\bar{y}_k}^2+{\bar{z}_k}^2} \\ 
	\frac{{\bar{x}_k} \dot{\bar{x}}_k + {\bar{y}_k} \dot{\bar{y}}_k}{\sqrt{{\bar{x}_k}^2+{\bar{y}_k}^2+		{\bar{z}_k}^2}} \\ 
	\tan^{-1}(\bar{x}_k, \bar{y}_k)
\end{array} \right].
\label{eqn:observation}
\end{align}
$r_k$ is the range, $\dot{r}_k$ is the range rate, $\theta_k$ is the azimuth angle, and $w_k \sim \mathcal{N}(0,\mathcal{R})$. The covariance matrix $\mathcal{R}$ is a diagonal matrix with the diagonal elements equal to the variances of the range, range rate, and azimuth angle measurements, which are denoted as $\sigma_{r_k}^2$, $\sigma_{\dot{r}_k}^2$, and $\sigma_{\theta_k}^2$ respectively. To compensate for the radar's platform motion, we define the coordinates $\bar{x}_k = x_k - x_k^P$ where $x_k^P$ is the $x$ coordinate of the sensor platform at time $k$; similarly for $\bar{y}_k$ and $\bar{z}_k$.

\subsection{SCFG and Syntactic Trajectory Modeling}
\label{subsec:scfgandtracking}

With the above model, we now show that if the modes $a_k \in \mathcal{T}$ in (\ref{eq:main}) are generated by a SCFG instead of a regular grammar, the target's
trajectory exhibits sophisticated geometric patterns.
For clarity, we focus on the following three examples
of geometric patterns: line, arc and m-rectangle (which is defined below). We  show below that  a line can be generated by a regular grammar,
but arcs and m-rectangles can be generated by SCFGs and cannot be generated by regular grammars.
Therefore, if we want to infer a target's intent by estimating whether it is moving in a line, arc or m-rectangle, we need to use SCFGs
and associated syntactic signal processing. To save space we will only describe rectangles and arcs that are aligned
with the horizontal and vertical axes. It is a trivial extension to consider rotated versions
of these trajectories. Similarly other trajectory patterns such as extended trapeziums, etc can be considered,
see \cite{Fu82} where complex patterns such as Chinese characters are considered.

{\em Language of Lines}: Recalling Definition \ref{def:scfg}, let  $\mathcal{L}_{\text{line}}$ denote the  language of  lines. It includes lines of arbitrary length, for example
the string $a^*$. Such strings can be generated by a regular grammar (Markov dependency).  For example, suppose we have a concatenated string $xA$,  where $x$ is any combination of nonterminals and terminals, and $A$ is a nonterminal, a one step derivation using the rule $A \rightarrow a A$ yields $xA \rightarrow xaA$. The derivation process is similar to that of a hidden Markov model.

{\em Language of Arcs}: 
The language of an arc, denoted $\mathcal{L}_{\text{arc}}$, can be compactly expressed as  $\mathcal{L}_{\text{arc}} =\{x \in a^n b^* c^n \}$, where there is same number of matching upward $a$ and downward $c$ modes and arbitrary number of forward modes $b$. For each $a$ in the string, there must be a matching $c$, and the corresponding grammar rule is $S \rightarrow aSc|\epsilon$, where $\epsilon$ is empty string. The arbitrary number of forward modes, on the other hand, can be modeled by the rule $S \rightarrow bS|Sb|\epsilon$. As a result, the basic production rules applied to construct arcs are $S \rightarrow aSc|bS|Sb|\epsilon$. However, as is known in the parsing literature, the inclusion of $\epsilon$ causes the parsing algorithm not to halt in all cases, $\epsilon$ is removed. The final equivalent production rules for an arc is $S \rightarrow aSc|bS|Sb|c$.

%The grammar can be constructed based on many techniques, and the two most commonly seen are matching and recursive relation.

The rules needed to generate patterns such as arc have syntax that is more complex than a regular grammar. Using the Pumping Lemma, we will show in Lemma 1 that a  HMM cannot model such an arc because of the self embedding (long range memory) -- the model needs to capture the fact that after $n$ steps in direction $a$, the target eventually moves by $n$  steps in the direction $c$. 
(Recall the definition of self-embedding  given in Sec. \ref{subsec:scfgbackground}).

{\em Language of m-Rectangles}: 
Let $\mathcal{L}_{\text{m-rectangle}}$ denote the language of m-rectangles (modified rectangles). Examples of m-rectangle strings are 
$h^n  b^+  d^n  f^+$, $h^+ d^n d^+ f^n$, etc. Thus a m-rectangle  is a 4 sided
geometrical pattern  
 comprising of  three left turns (or 3 right turns) each of ninety degrees, with
 two sides of equal length. Note that m-rectangles are not necessarily closed trajectories (if they were closed, they would coincide with a rectangle).
 
 Why do we consider m-rectangles  instead of rectangles? There are at least two reasons.
First, using
to the pumping lemma, Lemma \ref{lem:rect}   shows that the language comprising of rectangles is not a SCFG.
Second from a modeling point of view, in order to recognize suspicious behaviour of a target moving
around a building, m-rectangles are more robust since unlike a rectangle, the start and end points do not have to
coincide.

\begin{comment}
 Fig. \ref{GMTI:fig:track1} b) illustrates an instance of the m-rectangle language, and Fig. \ref{GMTI:fig:track1} c) demonstrates graphically of the derivation process listed below (full set of production rules for m-rectangle can be found in Sec. \ref{subsec:SCFGformulation}):
\begin{align*}
\text{m-rectangle} & \rightarrow \text{Bottom Down-Up Top Up-Down} \\
& \rightarrow \text{f Down-Up Top Up-Down} \\
& \rightarrow \text{f h Down-Up Top Up-Down} \\
& \rightarrow \text{f h h Top Up-Down} \\
& \rightarrow \vdots
\end{align*}
\end{comment}

{\bf Examples}:  To model the threat inference example provided at the beginning of Sec. \ref{sec:overview}, where a threat is related to suspicious U-turns and circling of a building, an arc language may be used to approximate U-turns and a m-rectangle language to circling around the restricted area.  The pincer operation, on the other hand, consists of two arcs in close proximity and of opposite direction. As a result, given continuous of the trajectories by the syntactic tracking,  a pincer operation can be identified by the following attributes: 1) two arcs of comparable size are identified, and 2) their locations are close together within a certain bound. Moreover, maritime events may also be identified by syntactic tracking. For example, a smuggling event may be modeled as one circling trajectory being approached by a linear trajectory. The labelling of trajectories can identify vessels that are loitering in the open sea, and detect other vessels moving toward them.

\subsection{Dynamics of Syntactic Motion Patterns as SCFG}
\label{subsec:SCFGformulation}

We are now ready to formulate the syntactic model for syntactic filtering using a SCFG. The kinematic modes of the multiple mode Bayesian filter, as illustrated in Fig. \ref{GMTI:fig:track1}a), are modeled by the terminal set
$$\mathcal{T} =  \{a,b,c,d,e,f,g,h\}= \{\pi/4, \pi/2, 3\pi/4, \pi, 5\pi/4, 3\pi/2, 7\pi/4, 2\pi\}. $$ The geometric patterns described in the previous section are modeled by the nonterminal set
\begin{equation}
 \mathcal{N}= \{L_a, L_b, \ldots, L_h, A_{ur},  A_{dr}, R_{cl}, R_{cc}, T_{cl}, T_{cc}, S\} .
 \end{equation}
The nonterminal $S$ is the starting symbol, and the meaning of the terminals and the nonterminals is explained below. Finally, the prior knowledge of the generation of the geometric languages in terms of the terminals and nonterminals is encoded by the production rules
\begin{align}
P = & \{ S \rightarrow  L_a | L_b | \ldots | L_h | A_{dr} | A_{ur} | R_{cl}|R_{cc}, \nonumber \\
& L_u \rightarrow  u~L_u | u \mbox{   for } u \in \mathcal{T}, \nonumber \\
& A_{ur} \rightarrow  aA_{ur} c|bA_{ur} |A_{ur}b|ac | b,  \nonumber\\
& A_{dr} \rightarrow  cA_{dr} a|bA_{ul} | A_{dr} b|ca | b, \nonumber \\
& R_{cl} \rightarrow T_{cl}~L_h, \nonumber \\
& T_{cl} \rightarrow b~T_{cl}~f | L_d, \nonumber \\ 
& R_{cc} \rightarrow T_{cc}~L_d, \nonumber \\
& T_{cc} \rightarrow b~T_{cc}~f | L_h \}. 
\label{eq:mrectprod}
\end{align}
\begin{comment}
& Sq_r \rightarrow TOP~UD~BOT~DU \nonumber \\
& Sq_l \rightarrow BOT~DU~TOP~UD \nonumber \\
& TOP \rightarrow L_b|L_f \nonumber \\
& BOT \rightarrow L_b|L_f \nonumber \\
& UD \rightarrow L_d \nonumber \\
& DU \rightarrow L_h
\end{comment}
The nonterminal $L_u$, $u \in \mathcal{T}$ generates lines in the direction $u$. $A_{ur}$ (respectively, $A_{dr})$ generates arcs pointing upward (downward) and to the right (see pincer in Fig. \ref{GMTI:fig:formation}). $R_{cl}$ and $R_{cc}$ are the clockwise and counter-clockwise m-rectangles respectively, and $T_{cl}$ and $T_{cc}$ are the turns that consist of the two equal length segments. The production rule of the turn $T$ and the arc $A$ are similar in form because they are both designed to capture the long range dependency of two line segments.
\begin{comment}
$Sq_r$ and $Sq_l$ are the clockwise and counter-clockwise m-rectangle respectively, and the nonterminals $TOP$, $UD$, $BOT$ and $DU$ represent the top, up-down, bottom, and down-up components of a m-rectangle. The  production rules of the m-rectangle presented include only one orientation, which can be identified by the sequence order of its four sides, but it is straightforward to include other orientations. The $TOP$ and $BOT$ are defined differently from $UD$ and $DU$ to capture the ambiguity between the clockwise and counter-clockwise m-rectangles. 
\end{comment}
It should be noted that the grammar is a small subset for illustrative purpose, and no intention is made to be exhaustive. The grammar is application specific, and it can be regarded as an guiding example for other development. The analysis of the grammar is provided in Sec. \ref{subsec:SCFGanalysis}.

Given the grammar, probability distribution is defined over the production rules. For each nonterminal $N$, the probability of its production rules must sum to 1, i.e. $$\sum_{\eta \in (\mathcal{N}\cup \mathcal{T})^* s.t. (A\rightarrow \eta)\in \mathcal{P}}  P(N \rightarrow \eta) = 1.$$ In practice, the production rule probabilities can be estimated from data.   The probability assignment has to follow a requirement to keep the grammar stable, and it will be discussed in the analysis that is presented in the next subsection.

\subsection{Structural Analysis of the SCFG Model}
\label{subsec:SCFGanalysis}

This section provides analysis of the languages presented in Sec. \ref{subsec:scfgandtracking}. Our results are the following: \\
(i) The relation $\mathcal{L}_\text{line}  \subset  \mathcal{L}_{RG} \text{ and } \mathcal{L}_\text{arc}, \mathcal{L}_\text{m-rectangle} \subset \mathcal{L}_{CFG}$ is  formally shown. More specifically, using the Pumping Lemma \cite{HMU07}, $\mathcal{L}_\text{arc}$ and $\mathcal{L}_\text{m-rectangle}$ are shown to be more general than regular grammars, and based on the structure of  their production rules, the languages are generated by CFGs, i.e.
$ \mathcal{L}_\text{arc}, \mathcal{L}_\text{m-rectangle}  \subset \mathcal{L}_{CFG} $.
 A regular grammar (HMM) cannot  generate exclusively  randomly sized m-rectangles
or only randomly sized arcs. (Of course a regular grammar can generate an arc or a m-rectangle
with some probability amongst a variety of random trajectories -- but that is of little use
in trajectory classification). It will also be shown that the language of rectangles is not CFG, which motivates the use of m-rectangles. \\
(ii) The second result provides conditions under which the SCFG model is well posed, and it boils down to checking the spectral radius of the stochastic mean matrix defined below.
 
 \subsubsection{Language of Trajectories}

The analysis of the geometric languages is based on the following Pumping Lemma that is proved in \cite{HMU07}.

\noindent  
\textbf{(i) Pumping Lemma for Regular Languages}:
Let $L$ be a regular language, then there exists a constant $K$ such that if $s$ is any string in $L$ such that $|s|$ is at least $K$ and for any way of breaking $s$ into $s=uvw$ with $|v| \geq K$, $v$ can be written as $xyz$ such that $y \neq \epsilon$ and $uxy^*zw \subseteq L$.

\noindent
\textbf{(ii) Pumping Lemma for Context-Free Languages}:
Let $L$ be a context free language, then there exists a constant $K$ such that if $s$ is any string in $L$ such that $|s|$ is at least $K$,  $s$ can be written as $s=uvwxy$, subject to the following conditions:
\begin{enumerate}
\item $|vwx| \leq K$. That is, the middle portion is not too long.
\item $vx \neq \epsilon$. Since $v$ and $x$ are the pieces to be "pumped", this condition says that at least one of the strings we pump must not be empty.
\item For all $i \geq 0$, $uv^iwx^iy$ in $L$. That is, the two strings $v$ and $x$ may be "pumped" any number of times, including $0$, and the resulting string will still be a member of $L$.
\end{enumerate}

Using the Pumping Lemma, we show that the arc and the m-rectangular languages are not regular.

\begin{lemma} \label{lem:arc} The arc trajectory language $\mathcal{L}_{arc}=\{a^n b^* c^n | n\geq 1\}$ is not regular.
\end{lemma}

\noindent
\textbf{Proof} Suppose $L$ is a regular language. Consider $s=a^Kc^K$, and choose $u=\epsilon, v = a^K$, and $w=c^K$. By the Pumping Lemma for regular languages, $s$ can be written as $s=uxyzw$ such that $y \neq \epsilon$ and $uxy^*zw \subseteq L$, which means for any $t \geq 0$, $uxy^tzw \in L$. When $t=0$, $a^{K-|y|}c^K \in L$. However, since $y \neq \epsilon$, $K-|y| < K$, and it contradicts the definition of $L$.

\begin{lemma} The m-rectangular trajectory language $\mathcal{L}_{m-rectangle}=\{a^n b^+ c^n d^+ | n\geq 1\}$ is not regular. \end{lemma}

\noindent
\textbf{Proof} Suppose $L$ is a regular language. Consider $s=a^Kbc^Kd$, and choose $u=\epsilon, v = a^K$, and $w=bc^Kd$. By the Pumping Lemma for regular languages, for any $t \geq 0$, $s$ can be written as $uxy^tzw \in L$. When $t=0$, $a^{K-|y|}bc^Kd \in L$. However, since $y \neq \epsilon$, $K-|y| < K$, and it contradicts the definition of $L$.

As mentioned in Sec.\ref{subsec:scfgandtracking},
we deal with  m-rectangles because the language generating standard rectangular trajectories is not context free. We now formally show this using the Pumping Lemma. The construction of a rectangular trajectory  can be expressed by a language $L=\{a^m b^n c^m d^n | m,n\geq 1\}$, where $m$ and $n$ signifies the length and width of the rectangle.  It is sufficient to show that a subset of the language, i.e. $L=\{a^n b^n c^n d^n | n\geq 1\}$ (which represents the language of square trajectories) is not context free.

\begin{lemma} \label{lem:rect} The rectangular trajectory language $L=\{a^n b^n c^n d^n | n\geq 1\}$ is not context free.\end{lemma}

\noindent
\textbf{Proof} Suppose $L$ is a context free language. Let $s=a^Kb^Kc^Kd^K$. The first condition dictates that $vwx$ is a substring of $a^Kb^K$ or $c^Kd^k$. Let $vwx$ be a substring of $a^Kb^K$, then $c^Kd^K$ is a substring of $y$, and $vx$ contains only $a$ and $b$. $uwy$ must be a string in the language by the Pumping Lemma, contains $K$ $c$'s and $d$'s, but has fewer than $K$ $a$'s and $b$'s. By contradiction, we can conclude that $L$ is not context free. Same steps can be applied when $vwx$ is a substring of $c^Kd^K$.
\vspace{1pt}

As a result, in order to deal with rectangular type trajectories in a CFG domain, m-rectangle language with the form $L=a^n b^+  c^n  d^+$ is considered.

\subsubsection{Well Posedness of the Model}
Before concluding this section, we need to address one more modeling issue. In a regular
grammar (HMM plus start and end states with non-zero probability of reaching the end state)
since there is no self-imbedding, the length of the data string generated is finite with probability one.
However, in a SCFG due to the self imbedding, it is possible for strings generated
by the production rules to never terminate.
Such instability is not desirable from a modeling point of view. So we need to restrict the model parameters to ensure that the generation of the geometric patterns is stable, i.e., the  derivation process is sub-critical \cite{MS92a}  and  terminates in finite time with finite length with probability one. This finiteness criteria provides a constraint on the SCFG model parameters, which may be used as a bound on the parameter values. We discuss this point by first defining the stochastic mean matrix.

\begin{definition} 
For $A,B \in \mathcal{N}$, the stochastic mean  matrix $M_{\mathcal{N}}$ is a $|\mathcal{N}| \times |\mathcal{N}|$ square matrix with its $(A,B)$th entry being the expected number of variables $B$ resulting from rewriting A:$$M_{\mathcal{N}}(A,B) = \sum_{\eta \in (\mathcal{N}\cup \mathcal{T})^* s.t. (A\rightarrow \eta)\in \mathcal{P}} P(A \rightarrow \eta)n(B;\eta).$$ Here $P(A \rightarrow \eta)$ is the probability of applying the production rule $A \rightarrow \eta$, and $n(B;\eta)$ is the number of instances of $B$ in $\eta$ \cite{Chi99}.
\end{definition}

The finiteness constraint is satisfied if the grammar satisfies the following theorem.

\begin{theorem} 
If the spectral radius of $M_{\mathcal{N}}$ is  less than one, the generation process of the stochastic context free grammar will terminate, and the derived sentence is finite.

\noindent
\textbf{Proof} The proof can be found in \cite{Chi99}.

\end{theorem}

\section{Syntactic Filtering Algorithms}
\label{sec:GMTIparsing}

Based on the SCFG modulated state space model constructed in Sec. \ref{sec:GMTImodeling}, algorithms to estimate the mode sequence and to perform the syntactic analysis are developed in this section. 
For example, we are interested in classifying whether the target trajectory is 
either a line, an arc or a m-rectangle.
Because the mode estimates are generated iteratively as the process unfolds, 
we use the Earley-Stolcke parsing algorithm to parse data from left to right recursively \cite{Sto95,IB00}. Earley-Stolcke parsing algorithm is a top down parser, and it is different from the more common bottom up parsers such as the CYK algorithm \cite{DEKM98}. 
Sec.\ref{subsec:scfgandtrackingb} gives an overview of the syntactic parsing approach.
Sec. \ref{subsec:trackers} discusses the implementation of the mode estimator that produces estimates of mode sequences, and Sec. \ref{subsec:earleyparser} summarizes the implementation of the syntactic pattern estimator based on the extended version of the Earley-Stolcke parser.

\subsection{Syntactic Parsing and Target Tracking}
\label{subsec:scfgandtrackingb}

 The operation of inferring the production rules used given a string of terminals (e.g. fhhbd) is called  stochastic parsing, and in the context of syntactic filtering, given a SCFG, a track consists of both a sequence of kinematic estimates and a set of parser states. The definition of a parser state and its semantics in terms of a track in target tracking are discussed in this section, and the algorithm that recursively computes parser states from kinematic measurements is presented in Sec. \ref{subsec:earleyparser}.

The  Earley Stolcke parser described below can be viewed as a  generalization of the forward algorithm (which is used for HMMs) to the SCFG \cite{Sto95}. Given the string
of terminals $a_{1:N}$ from the tracker, the control structure the parser uses to store incomplete parse trees is defined as 
\begin{equation}
k: X_i \rightarrow \lambda . Y \mu [\alpha, \gamma],
\label{eqn:parserstate}
\end{equation}
where $X$ and $Y$ are nonterminals, $\lambda$ and $\mu$ are substrings of nonterminals and terminals, and $\lambda$ contains the string $a_{i:k}$. "." is the marker that specifies the end position, indexed by $k$,  and $i$ is the beginning index of the substring that is partially parsed by the nonterminal $X$. $\alpha$ is called forward probability and it is the sum of probabilities of all incomplete parse trees containing $a_{1:k}$, and $\gamma$ is called inner probability and it is the sum of probabilities of all incomplete parse trees containing $a_{i:k}$. 

\begin{figure}
\centering
\psfrag{a1}{$a_1$}
\psfrag{a2}{$a_2$}
\psfrag{ai}{$a_i$}
\psfrag{ak}{$a_k$}
\psfrag{aN}{$a_N$}
\includegraphics[width=0.65\linewidth]{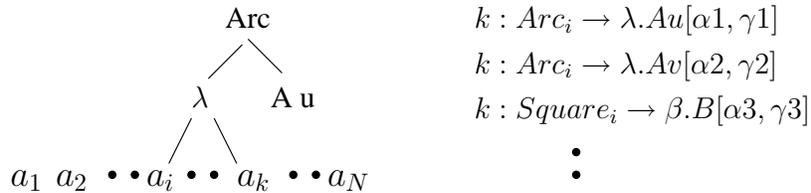}
\caption[Figure illustrates an example Earley parser state]{Syntactic analysis parses the mode sequence by dynamically creating valid parser states for each $i$ and $k$. Each parser state can be view as a hypothesis whose likelihood is indicated by $\alpha$ and $\gamma$. The figure illustrates few parser states and also the pictorial representation of a particular parser state, $k:Square_i \rightarrow \lambda.A u$, in terms of the mode sequence it represents.}
\label{GMTI:fig:parsingEg}
\end{figure}

Illustration of syntactic analysis for syntactic filtering is provided in Fig. \ref{GMTI:fig:parsingEg}. 
Consider a trajectory generated by (\ref{GMTI:eqn:statespace}) and a mode sequence $a_{1:k}$ that is estimated as  a string of terminals from the trajectory. At each time $k$, $a_k \in \mathcal{T}$ denotes the target's kinematic mode, i.e., its direction of travel, the aim of syntactic analysis is to infer the geometric patterns that might have produced the trajectory based on a SCFG formulation.  Syntactic analysis recursively builds different parse trees, represented by a collection of parser states, as hypotheses to "explain" the geometric patterns. (Details are provided in Sec. \ref{subsec:earleyparser}.) More specifically, syntactic filtering extends multiple mode tracking algorithm with the incorporation of syntactic analysis, and the semantics of the parser state (\ref{eqn:parserstate}) are summarized here:
\begin{itemize}
\item Radar scans $i$ to $k$ are processed by the parser, and the position of the current scan $k$ in the input mode sequence is labeled by the dot ".".
\item Nonterminal $X$ represents a geometric pattern and it is a hypothesis used to characterize the input mode sequence generated by scans $i$ to $k$.
\item $\alpha$ keeps the likelihood probability of the mode sequence $a_{1:k}$ given the nonterminal, and $\gamma$ the likelihood probability of $a_{i:k}$.
\item Future mode evolution could be predicted based the production rules of $Y$.
\end{itemize}
In other words, syntactic filtering tracks the evolution of the mode sequence, and iteratively builds different hypothesis trees of nonterminals (geometric patterns and their elements) to explain the mode sequence. 

%More specifically, 
%\begin{equation}
%P(m|a_{1:k},G_m) = \frac{P(a_{1:k}|m,G_m)P(m|G_m)}{P(a_{1:k}|G_m)},
%\label{GMTI:eqn:syntactic}
%\end{equation}
%and $$P(a_{1:k}|m,G_m) = \sum_{t=\mbox{parse}(a_{1:k})} P(t|m,G_m)$$ is the probability of the parse trees that generate the mode sequence $a_{1:k}$ given the grammar $G_m$. 

\subsection{Syntactic Enhanced Tracker}
\label{subsec:trackers}

The mode estimator (\ref{GMTI:eqn:meta}) that computes $a^*_{1:k}$ can be implemented using any approximate multiple mode Bayesian tracker, for example, an  extended Kalman with IMM or a multiple mode particle filter.  In either case,  the nonlinearity in the observation model implies that an approximate filter needs
to be used since  finite-dimensional optimal filters do not exist.  As will be described  below, 
the multiple mode tracker  outputs  the mode probability $w_k^j$ for mode $j$. It is this mode probability
estimate that is fed into the syntactic parser described in Sec. \ref{subsec:earleyparser} .

\subsubsection[Multiple Model Sequential MCMC (particle filter)]{Multiple Model Sequential Markov Chain Monte Carlo (particle filter)}

 Let $y_k = (x_k^\prime, a_k)^\prime$, where $x_k$ is a continuous value kinematic state, $a_k$ is a discrete value IMM mode, and $^\prime$ denotes transpose. The posterior probability distribution of the state space is approximated by $P(y_k|z_k) = \sum_{i=1}^N w_k^i \delta(y_k - y_k^i)$. The random measure $\{y_k^i, w_k^i\}_{i=1}^N$ are the particles and their associated weights to characterize the posterior distribution, and $N$ is the number of particles. The multiple mode particle filter algorithm consists of three steps \cite{RAG04}: 
\begin{enumerate}
\item sampling of the IMM mode transitions, 
\item sampling of the mode conditioned kinematic state, and 
\item resampling to avoid degeneracy. 
\end{enumerate}
These three steps are now described:

Given the set of IMM modes $\{ a_{k-1}^i\}_{i=1}^N$ at time $k-1$, the sampling of the IMM mode 
involves generating $\{a_k^i\}_{i=1}^N$ based on the transition matrix $\pi_{ij}$.
%plemented based on the inverse transform method. generate  based on the its transition matrix. Let $C_{a_{k-1}}(a)  = \sum_{j=1}^a be the cumulative distribution of the IMM mode $a$ given $a_{k-1}=i$, the algorithm to generate the modes is shown below:
%\begin{algorithmic}
%\FOR {$i$ = 1 to $N$}
%\STATE Sample $u \sim U[0,1]$
%\STATE $m \gets 1$
%\WHILE {$C_{a_{k-1}^i}(m) < u$}
%\STATE $m \gets m+1$
%\ENDWHILE
%\STATE $a_k^i \gets m$
%\ENDFOR
%\end{algorithmic}

The sampling of the mode conditioned kinematic state involves sampling from the transition probability and calculating the associated weight. The optimal importance density is $P(x_k|x_{k-1}^i,a_k^i,z_k)$ given the IMM mode sampled from step 1, yet the most popular and simpler importance function is $P(x_k|x_{k-1}^i,a_k^i)$. The un-normalized weight of each sampled particle is updated by the following equation
$$\tilde{w}_k^i = w_{k-1}^i \frac{P(z_k|x_k^i,a_k^i)P(x_k^i|x_{k-1}^i,a_k^i)}{q(x_k^i|x_{k-1}^i,a_k^i,z_k)},$$
where $q(y_k^i|y_{k-1}^i,z_k)$ is the importance density. Using the simplified importance density, it becomes $$\tilde{w}_k^i = w_{k-1}^i P(z_k|x_k^i,a_k^i).$$ The normalized weight is then $w_k^i = \tilde{w}_k^i / \sum_{i=1}^N \tilde{w}_k^i$.

The resampling involves a mapping of random measure $\{x_k^i, w_k^i\}$ to $\{x_k^{i*}, 1/N\}$ with uniform weights. The resampled particles $\{{x}_k^{i*}\}_{i=1}^N$ are generated by resampling with replacement $N$ times from the random measure $\{x_k^i, w_k^i\}_{i=1}^N$. The resampling is necessary if the effective sample size is less than a threshold sample size, and the effective sample size is computed as
$$\hat{N}_{eff} = \frac{1}{\sum_{i=1}^N (w_k^i)^2}.$$
If resampling is not performed, degeneracy problem would occur which means after a certain recursive steps, all but one particle will have negligible normalized weights.

\subsubsection{Extended Kalman filter with IMM} \label{ekfimm}

Because Eq. (\ref{eqn:observation}) is highly nonlinear, extended Kalman filter is needed to process the observations. Consider the following measurement model:
 $$\tilde{z}_k=\tilde{h}(x_k) + \tilde{w}_k$$
where
\begin{equation}
\tilde{h}(x_k) = \left[ \begin{array}{c} r_k \sin\theta_k \\ r_k \cos\theta_k \\ \dot{r}_k \end{array} \right] 
= \left[ \begin{array}{c} x_k \\ y_k \\ \dot{r}_k \end{array} \right]
\end{equation}
and $\tilde{w}_k \sim N(0,\tilde{R})$ is the measurement noise in the converted model. The converted covariance matrix is 
\begin{align*}
&\tilde{R}  = \left( \begin{array}{ccc}
\sigma_x^2 & \sigma_{xy} & 0 \\
\sigma_{yx} & \sigma_y^2 & 0 \\
0 & 0 & \sigma_{\dot{r}_k}^2
\end{array} \right),
\end{align*}
whose elements are
\begin{align*}
\sigma_x^2  = & r_k^2 \sigma_{\theta_k}^2 \cos^2\theta_k + \sigma_{r_k}^2 \sin^2\theta_k \\
\sigma_{xy} = & (\sigma^2_{r_k}-r_k^2\sigma_{\theta_k}^2)\sin\theta_k \cos\theta_k \\
\sigma_y^2 = & r_k^2\sigma_{\theta_k}^2 \sin^2\theta_k + \sigma_{r_k}^2\cos^2 \theta_k.
\end{align*}

In order to run extended Kalman filter, the Jacobian of the converted measurement function is 
$$\nabla_{x_k} \tilde{h}(x_k)=
\left(
\begin{array}{cccc}
1 & 0 & 0 & 0 \\
0 & 1 & 0 & 0 \\
\frac{\partial \tilde{h}[3]}{\partial x_k} & \frac{\partial \tilde{h}[3]}{\partial y_k} & \frac{\partial \tilde{h}[3]}{\partial \dot{x}_k} & \frac{\partial \tilde{h}[3]}{\partial \dot{y}_k}
\end{array}
\right)
$$ 

As will be shown in Sec. \ref{sec:GMTIparsing}, the terminal probability $w_k^j = P(a_k = j|z_{1:k})$ models the input uncertainty for the parsing process, and the position estimate $\hat{\mathbf{x}}_{k|k}$ is stored in the low and high marks of the Earley state for enforcing consistency of the tracks. According to the kinematic model, we can compute the two variables based on the interacting multiple models (IMM) \cite{KBP00}, and its algorithm is summarized here:

\begin{itemize}
\item Calculating the mixing probabilities
\begin{align*}
u_{k-1}^{i|j} & = P(a_{k-1}(i)|a_k(j),z_{1:k-1})  \\
& = \frac{1}{c} P(a_k(j)|a_{k-1}(i),z_{1:k-1})P(a_{k-1}(i)|z_{k-1})
\end{align*}
\item Mixing
\begin{align*}
\hat{x}_{k-1|k-1}^j = & \sum_{i=1}^8 u_{k-1}^{i|j} \hat{x}_{k-1|k-1}^i \\
P_{k-1|k-1}^j = & \sum_{i=1}^8 u_{k-1}^{i|j}\left[ P_{k-1|k-1}^i + ( \hat{x}_{k-1|k-1}^i - \hat{x}_{k-1|k-1}^j) \right. \\
& \left. (\hat{x}_{k-1|k-1}^i - \hat{x}_{k-1|k-1}^j)^\prime\right] 
\end{align*}
\item Model-matched filtering
$$\Lambda_k^j = p(z_k|z_{1:k-1},a_k=j)$$
\item Mode probability update
%\begin{align*}
$$w_k^j =  \frac{\Lambda_k^j \sum_{i=1}^8 \pi_{ij} w_{k-1}^i}{\sum_{j=1}^8 \Lambda_k^j \sum_{i=1}^8 \pi_{ij} w_{k-1}^i} $$
%u_{k-1}^{i|j} = & P(t_{k-1}=i|a_k=j, z_{1:k-1}) = \frac{\pi_{ij} w_{k-1}^i}{\sum_{i=1}^8 \pi_{ij} w_{k-1}^i}
%\end{align*}
\item Estimate and covariance combination
\begin{align*}
\hat{x}_{k|k} = & \sum_{j=1}^8 \hat{x}_{k|k}^j w_{k}^j \\
P_{k|k} = & \sum_{j=1}^8 w_k^j \left[ P_{k|k}^j + [\hat{x}_{k|k}^j - \hat{x}_{k|k}][\hat{x}_{k|k}^j - \hat{x}_{k|k}]' \right]
\end{align*}
\end{itemize}

\subsection{Extended Earley Stolcke Parsing of Target Trajectory}
\label{subsec:earleyparser}

%Extension are for integration of parsing and tracking. Expand on it!!

We are now ready to describe the syntactic signal processing algorithms with Earley Stolcke parser, and also the extensions of the parser needed to integrate it with the tracking algorithm described above. Recall the system framework illustrated in Fig. \ref{GMTI:fig:framework}, the parser assumes the existence of tracking and data association modules, and performs syntactic analysis of their outputs. The parser is extended to 1) model the uncertainties of the mode estimates generated by the Bayesian tracker, 2) keep parsing robust against non-detections generated by the data association module, 3) perform track initiation for syntactic filtering, and 4) prune unlikely tracks to trade-off track completeness with lower computational complexity. The extensions are largely based on those described in \cite{IB00}, but altered to fit the specific case of syntactic filtering with GMTI measurements. The extensions are discussed later when parsing operations are introduced.

In order to introduce the extensions, modifications to both the parser state and the production rules are necessary. The parser state of the Earley Stolcke parser is redefined as
$$k: X_i \rightarrow \lambda . Y \mu [l, h, \alpha, \gamma],$$
where $l$ is the kinematic state of the track at scan $i$ and $h$ the state at scan $k$. Let $d$ be the euclidean distance, and $f(d)$ a similarity function to measure the spatial correlation of two kinematic states. Many spatial correlation models may be applied \cite{BOS01}, and the function used in this paper is a power exponential function, $f(d) = \exp(-(\frac{d}{\theta_1})^{\theta_2})$, where $\theta_1 > 0$ and $\theta_2 \in (0,2]$ are determined experimentally. The production rule, on the other hand, is modified to model non-detection events due to both a miss or target moving slower than the minimum detectable velocity. For every production rule that involves the generation of terminals, a nonterminal $N_d$ is added, i.e. the rule $L \rightarrow l L$ will be modified to include $L \rightarrow l L | N_d L$, where $N_d$ will be mapped to a non-detection returned by the data association module.

\vspace{3pt}
\noindent
{\em Parsing Example:}
To give more intuition, here is a simple example of parsing a very short input string ``bb".
The steps are illustrated in Table \ref{table:bb}. For simplicity, only a subset of the production rules listed in (\ref{eq:mrectprod}) are used, only the line terminals, i.e. $L_a, L_b, \ldots, L_h$, and their associated production rules are used.

To initialize the parsing process, a dummy parser state $0:{}_0 \rightarrow .S [l_c, h_c, 1,1]$ is inserted, where $l_c$ and $l_h$ are the extracted kinematic states of the target from the GMTI detection. The dummy parser state is the first entry in column $0$ of the table, and it indicates that at the index position 0, the start symbol is applicable to parse the input string. With the dummy parser state in place, the parser builds the parse tree by iteratively applying three operations:  \textit{prediction, scanning, and completion,} which will be discussed in detail later. The operations are applied sequentially, and each operation works on the set of parser states produced by the previous operation.

Given a set of parser states (which contains only the initial dummy parser state at index 0), the prediction operation searches for parser states whose index marker has a nonterminal to its right. (In the case of the dummy parser state, the nonterminal to the right of the index marker is the start symbol $S$). For those nonterminals, the prediction operation generates a set of predicted states with their production rules. Please see the entries below the dummy parser state under the heading ``Prediction". Given the predicted parser states, the scanning operator looks if there are parser states whose index marker has a terminal to its right. If the terminal of those parser states matches the input string at the indexed position, their index markers are advanced by one position. The generated parser states are called the scanned parser states. Please see the entries in column 1 under the heading ``Scanning". It can be seen only the predicted parser states with terminal $b$ are advanced because the input terminal at index 1 is $b$. Lastly, given the scanned parser states, the completion operation looks if there are parser states whose index marker is at the end of its production rule. If any are found, the parser states that generated those scanned parser states will have their index advanced by one position. Please see the entries under the heading ``Completion in column 1. The completed parser state $1:L_b {}_0 \rightarrow b.$ generates the completed state $1:S_0 \rightarrow L_b .$. The three operations will be applied iterative until the dummy state is completed. The details of the three operations are discussed next in turn.

\begin{table}
\centering
\begin{tabular}{lll}
0						& 1							& 2 \\
\hline
						& $b$						& $b$ \\
\hline
$0: {}_0 \rightarrow .S$ 		& Scanning 					& Scanning \\
Prediction 				& $1: L_b {}_0 \rightarrow b. L_b$ 	& $2: L_b{}_1 \rightarrow b. L_b$ \\
$0: S_0 \rightarrow .L_a$ 		& $1: L_b {}_0 \rightarrow b.$ 		& $2:L_b {}_1 \rightarrow b.$ \\
$0:S_0 \rightarrow .L_h$	 	& Completion 					& Completion \\
$0:L_a {}_0 \rightarrow .a L_a$& $1: S_0 \rightarrow L_b . $		& $2:L_b {}_0 \rightarrow b L_b.$ \\
$0:L_a {}_0 \rightarrow .a$	& Prediction				 	& $2:S_0 \rightarrow L_b.$ \\ 
$\hdots$					 & $1: L_b {}_1 \rightarrow .b L_b$  	&  \\
$0:L_h {}_0 \rightarrow .h L_h$	& $1: L_b{}_1 \rightarrow .b$		& \\
$0:L_h {}_0 \rightarrow .h$	&							& \\
\hline
\end{tabular}
\caption[Simplified example demonstrating the Earley Stolcke parsing algorithm]{Earley Stolcke Parser parsing a simple terminal string "bb" with the simplified grammar specified in Sec. \ref{subsec:SCFGformulation}; only the production rules associated with the nonterminal Line are included.}
\label{table:bb}
\end{table}

\subsubsection{Prediction}
The prediction operator adds parser states that are applicable to explain the unparsed input string. For all parser states of the form 
$$k : X_i \rightarrow \lambda .Y \mu~[l,h,\alpha,\gamma],$$
where $\lambda$  and $u$ may be empty, and $Y$ is the nonterminal, the operator adds 
$Y$'s production rule, 
$$k : Y_k \rightarrow .v~[l,h,\alpha', \gamma'],$$ 
as a predicted parser state. The $\alpha'$ and $\gamma'$ are updated according to 
$$\alpha' = \sum_{\lambda,u} \alpha(k : X_i \rightarrow \lambda.Z\mu)\mathcal{R}_L(Z,Y)P(Y\rightarrow v)$$ and 
$$\gamma' = P(Y \rightarrow v),$$ 
where $\mathcal{R}_L$ is a reflective transitive closure of a left corner relation and it computes the probability of indefinite left recursion in the productions. (The detail of the relation is omitted as it has little significance in this paper. Interested readers can refer to \cite{Sto95}.) The new predicted parser state inherits the kinematic states because it explains the same substring of the mode sequence. The pruning capability of the parser can be implemented by discarding the predicted parser states if its forward probability is lower than a threshold. The value of the threshold balances system loading and track completeness. In addition, the prediction stage may also be modified to capture a track with an unknown beginning. At each time instant when the prediction operation is run, a dummy parser state of the form $\forall k ~~ k:{}_k \rightarrow .S$ can be inserted if there are GMTI detection that cannot be associated with any partial parse tree. With this dummy state, the parser is not limited to capture patterns that were started at the time instant 0.

\subsubsection{Scanning}
The scanning operator matches the terminal in the input string to the parser states generated from the prediction operator. For all parser states of the form 
$$k : X_i \rightarrow \lambda.a\mu~[l,h,\alpha,\gamma],$$ 
where $\lambda$ and $\mu$ can be empty, the parser state 
$$k+1:X_i \rightarrow \lambda a.\mu~[l,\mathbf{x}_a,\alpha',\gamma']$$ 
is added if the terminal at $k+1$ is $a$, where $\mathbf{x}_a$ is the kinematic state of the terminal $a$ estimated by the Bayesian filter, and $P(a)$ is its probability distribution (uncertainty of the mode estimate from the Bayesian filter). The $\alpha'$ and $\gamma'$ are updated according to 
$$\alpha' = \alpha(k : X_i \rightarrow \lambda.a\mu)P(a)$$ and 
$$\gamma' = \gamma(k : X_i \rightarrow \lambda.a\mu)P(a).$$ 
It is noted that by including $P(a)$ in updating $\alpha$ and $\gamma$, the parsing process also takes the input uncertainty in account.

\subsubsection{Completion}
The completion operator advances the marker position of the pending predicted parser states if their derived parser states match the input string completely. The scanned parser states whose marker is at the end of their rule have the form 
$$k : Y_j \rightarrow v.~[l_2,h_2,\alpha'',\gamma''],$$ 
and it has corresponding parser states (pending predicted parser states) of the form 
$$j : X_i \rightarrow \lambda.Y\mu~[l_1,h_1,\alpha,\gamma],$$
i.e. the parser states that generated the scanned parser states at the prediction stage. The two parser states generate and add a completed parser state 
$$k : X_i \rightarrow \lambda Y.\mu~[l_1,h_2,\alpha',\gamma'].$$ 
It is important to notice how the indices of the parser states are related. The indices of the pending predicted parser state indicate that the nonterminal $X$ was applied at $i$, and its derived parser state (the scanned parser state) indicates that $Y$, which corresponds to a substring of $X$, matches the terminal substring $j$ to $k$, it can then be concluded that the pending predicted parser state can now explains the substring $i$ to $k$ so its marker is advanced accordingly. The associated $\alpha$ and $\gamma$ probabilities are updated according to
$$\alpha'=f(h_1,h_2)\sum_v \alpha(j : X_i \rightarrow \lambda.Z\mu)\mathcal{R}_U(Z,Y)\gamma''(k : Y_j \rightarrow v.)$$ and 
$$\gamma'=f(h_1,h_2)\sum_v \gamma(j : X_i \rightarrow \lambda.Y\mu)\mathcal{R}_U(Z,Y)\gamma''(k : Y_j \rightarrow v.)$$ respectively,
where $\mathcal{R}_U$ is a reflective transitive closure of a unit production relation and it computes the probability of an infinite summation due to cyclic completions (interested reader can refer to \cite{Sto95} for more detail), and the similarity function here models the consistency between the pending predicted parser state and the completed parser state. If the likelihood probabilities of the completed parser state is lower than a threshold, it will be pruned to trade track completeness with computation reduction. 

The parsing algorithm can be  extended to incorporate further  domain knowledge
of the human operator. For example, selection logic can be added to the prediction operator, that instead of adding all probable states, only adds those whose production rules yield terminal symbols  compatible with the input string. In other words, instead of purely top down parsing, bottom up information could be incorporated to speed up the parsing algorithm. 

%\subsubsection{Spatial Extension of the Parser}

%A parser is instantiated for each target observed, and the parsing states generated are stored in a central database where another parser is at work. Two levels of parsing are in place; an geometric parser for each target, and a situation awareness parser parsing through the geometric parsing states generated.

%For example... a convoy consists of several objects in a row.... the spatial parsing in Fu's book.

%Two possible implementations:
%1) Fusion of parsing states from multiple parsers in a table enables spatial filtering to be performed and search for situations such as convoys and pincer.

%2) Single parser that parse all detections and store all partial trees in a single table.

%It's important to label each parsing state with time and position information so the parsing states can be combined easily, and spatial filtering possible.

\section{Experimental Setup and Results}
\label{sec:simulation}

The numerical studies in this section demonstrate how stochastic parsing with target tracking can discern geometric patterns with real GMTI data collected by DRDC. Sec. \ref{subsec:experiment} describes the experiment setup and the data model. Sec. \ref{subsec:preprocess} discusses the pre-processing required to transform measurements from various coordinate systems. Sec. \ref{subsec:numerical} summarizes the numerical results. Finally, Sec. \ref{GMTI:subsec:feedback} shows that by  feeding back the higher level syntactic estimates to the standard tracker, substantial improvements in performance are possible.

\subsection{Experimental Setup}
\label{subsec:experiment}

\begin{figure}
	\begin{center}
		\includegraphics[scale=0.3]{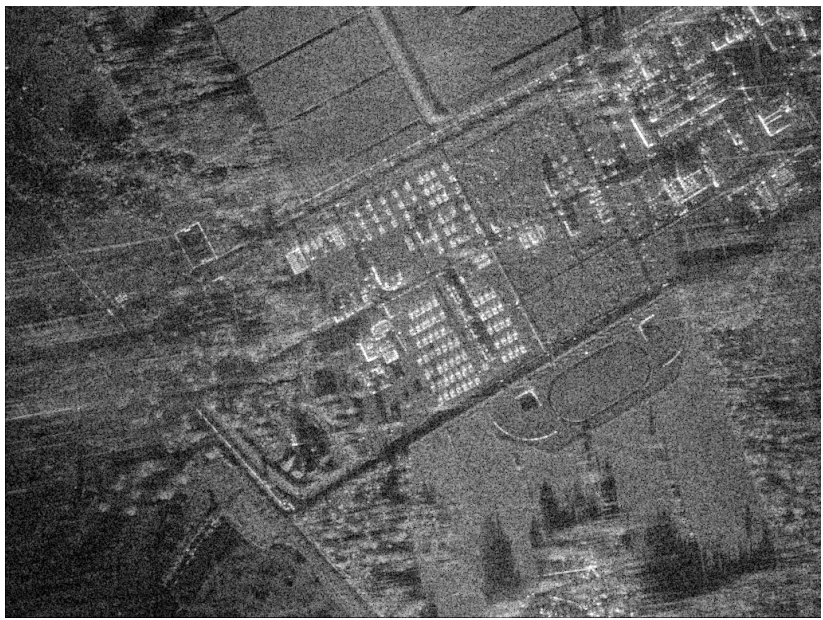}
	\end{center}
	\caption[SAR image captured by DRDC XWEAR system]{A SAR image of the location of the experiment captured by the DRDC XWEAR system.}
	\label{GMTI:fig:SARImageofCR}
\end{figure}

The GMTI data is collected using DRDC Ottawa's X-band Wideband Experimental Airborne Radar (XWEAR)\cite{DMH03, BD06}. It is a reflector-antenna-based multi-function radar that is designed to collect coherent radar echos with various modes for wide area search and imaging. The XWEAR radar's data collection modes include search modes, where the antenna is rotating, stripmap SAR and spotlight SAR imaging modes, and wide-area surveillance GMTI mode. The introduction of a multimode feed, i.e., the ability to carry two electromagnetic modes, enables a two-channel GMTI capability\cite{BD06}. The XWEAR radar is used to collect data for investigations into wideband synthetic aperture radar (SAR), inverse SAR (ISAR), maritime surveillance, and GMTI.  

The navigation subsystem of the XWEAR radar consists of an inertial measurement unit (IMU) mounted near the antenna phase centre (APC), and an embedded global positioning/inertial navigation system (EGI) mounted near the centre of gravity of the aircraft. In order to collect coherent radar echoes, the radar data needs to be compensated for undesirable APC motion (e.g., changes in aircraft ground speed and deviation from ideal flight path) that introduces pulse-to-pulse errors. The IMU provides high-rate (200 Hz) measurements of velocity and angular increments.   The strap-down navigator algorithms process these measurements and yield estimates of APC position and velocity, and antenna orientation. The EGI blends its own inertial data with  GPS data using an internal Kalman filter and the resulting accuracy in position and velocity is about 2 m and 0.03 m/s respectively. The EGI output is used in an external Kalman filter to give long-term stability to the strap-down navigation solution from the IMU. The phase corrections are then applied relative to a reference trajectory, so that the resulting data is coherent. 

In flight trials, the radar was installed and flown on a Convair 580 aircraft. The data was collected over western Ottawa. A SAR image of the scene is shown in Figure \ref{GMTI:fig:SARImageofCR}. The aircraft was moving at about 200 knots, or 100 m/s, with aircraft positions recorded as discussed above. The ground moving target is a truck that is moving in trajectories that form various geometric patterns.  The GPS data of the truck was also recorded for ground truth. The antenna was pointed to a fixed point on the ground, and the target always had non-zero radial velocity so that the target could be observed continuously by STAP-based GMTI techniques.  %The truck moved along the road in the top half, road parallel to it, and the roads perpendicular to it and at the far right corner of the figure. 
 The elevation angle is neglected as it does not provide any additional information. This is because in the GMTI case, the target is moving on a known plane. Then, if the pointing angle and range resolution are known, a particular range bin is equivalent to an elevation angle of the target. 

\begin{table}
	\centering
	\begin{tabular}{|c|c|}
		\hline
		Pulse Length& 5$\mu$s\\
		PRF&1-2 kHz\\
		Carrier Frequency&9.75 GHz\\
		Polarization&Transmit and Receive-Horizontal\\
		Antenna &1m width, 2.5$^o$ (4$^o$) azimuth (elevation) beamwidth\\
		\hline
	\end{tabular}
	\caption{Radar parameters of the DRDC XWEAR system used in data collection.}
	\label{tab:XWEAR}
\end{table}

%GMTI has range resolution of 1 to 10 meters
%and it takes 0.1 second for GMTI to produce a cluster suppressed frame.
%CPI = 128
%GMTI/STAP  XWEAR DRDC radar
%Stripmap/SAR
%Spotlight/SAR
%2 channel radar (1 transmitter and 2 receiver)
%0.1 second gives coarse image
%1 second gives high resolution image
%5 to 10 seconds give higher resolution image

\subsection{GMTI Dataset}
\label{subsec:preprocess}

Detection using STAP was carried out using a coherent processing interval (CPI) of about 128 pulses and the pulse repetition frequency was 1 kHz. The duration of the data acquisition studied here is about 108 seconds. Since the target of interest had a fairly high SNR and moved above the minimum detectable velocity of the GMTI sensor for a significant fraction of the time, move-stop-move pattern is not considered in this instance. In addition, the tracker was not fed all of the detections that were found at every CPI as there were several false alarms. Instead, only detections that were present in 3 (or more) out of 7 consecutive CPIs were used in the tracking algorithm.

Since tracker inputs are based on several CPIs, a target need not be detectable at every CPI. Similarly, by requiring multiple detections in a set of CPIs, several false alarms could be eliminated. This was found to be sufficient to eliminate false alarms for this data set, although a more sophisticated tracking algorithm will be required for targets that have low SNR. 
%The use of a longer measurement interval (about 0.9 sec, rather than 0.128 sec) for tracking was found to yield acceptable tracking performance for this data set.
The standard deviations used in the GMTI measurement model for range, azimuth angle, and range rate, were 5 m, 2.5 degrees, and 0.1 m/s respectively, and the state model noise used for the CV model was chosen to be 0.05 and 0.5 for the parallel and the orthogonal component respectively. No terrain data is used to modulate the measurement model.

The sensor platform coordinates, provided by the global positioning system on-board the aircraft, are given in the geodetic coordinate system. The GMTI measurements, which include range, range rate, and azimuth angle, are collected in the local spherical coordinates. The tracking algorithms developed are defined in a tangential plane Cartesian coordinate system. As a result, in order to apply the tracking algorithms developed, it is necessary to express the GMTI measurements in terms of quantities defined on the tangential plane Cartesian coordinates. The origin of the Cartesian coordinates is chosen to be the ECEF coordinates of the scene centre.

\begin{comment}
The tracking algorithms developed are defined in a tangential plane Cartesian coordinate system. The sensor platform coordinates, provided by the global positioning system on-board the aircraft, are given in the geodetic coordinate system. The GMTI measurements, which include range, range rate, and azimuth angle, are collected in the local spherical coordinates. As a result, in order to apply the tracking algorithms developed, it is necessary to express the GMTI measurements in terms of quantities defined on the tangential plane Cartesian coordinates. The process used is summarized next.

The origin of the Cartesian coordinates is chosen to be the ECEF coordinates of the first estimated target location based on the first detection. The APC locations at the two instants are converted to the ECEF coordinates. The z-axis is chosen to be given by the unit vector formed from the vector difference (in ECEF coordinates) of the APC and a point directly below on the ground at time of the first measurement. The x-axis is defined by the vector difference of the second APC location and the first APC location. Finally, the y-axis is defined as the cross product of the x and the z axes. Given the x, y, and z axes, all the quantities can be transformed to Cartesian coordinates by suitable dot product with the axes.
\end{comment}

\subsection{Numerical Studies of Syntactic Filtering}
\label{subsec:numerical}

\begin{figure}
%\begin{minipage}{0.5\linewidth}
\includegraphics[width=0.8\linewidth]{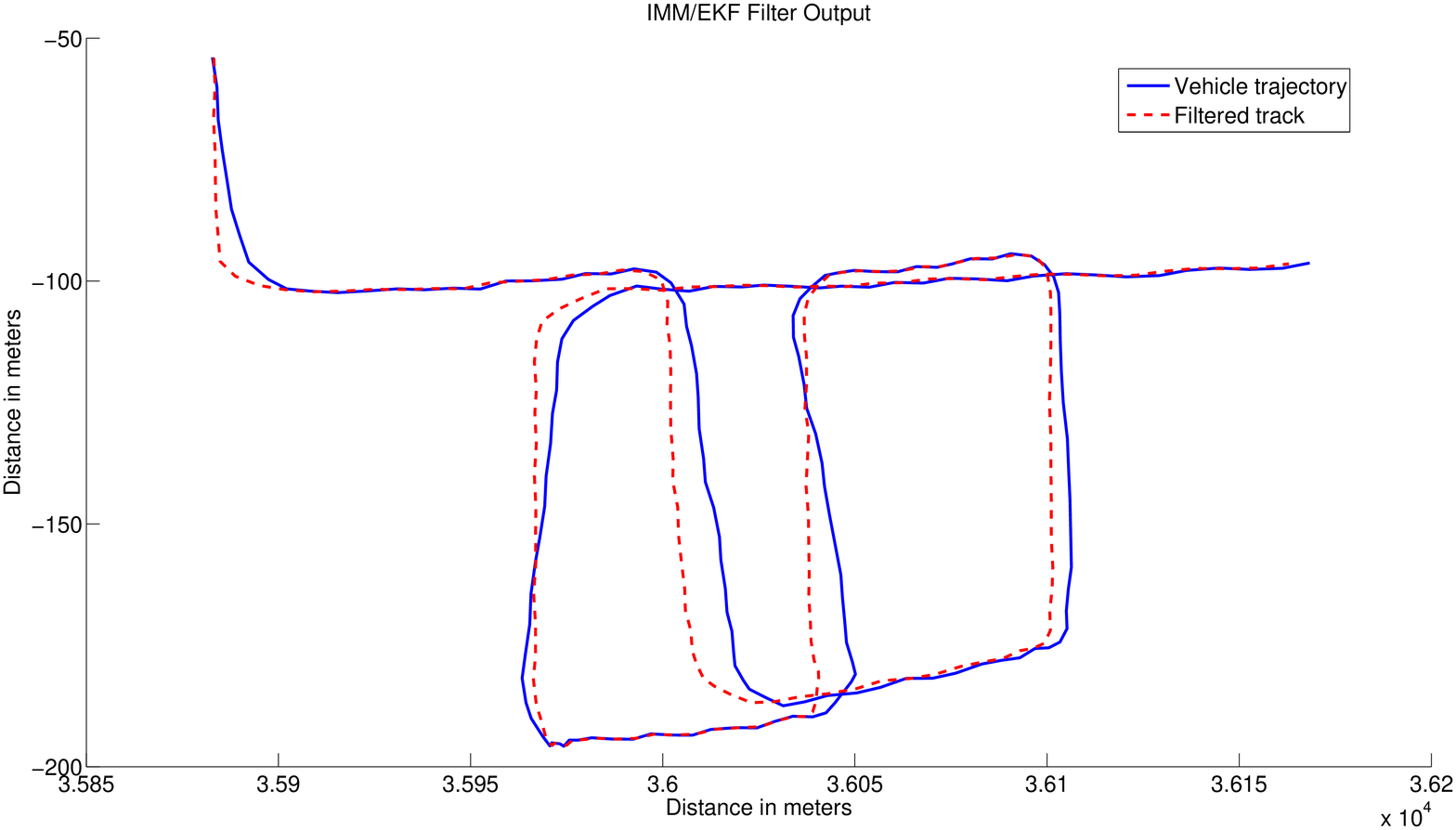} \\
%\end{minipage}
%\begin{minipage}{0.45\linewidth}
\includegraphics[width=0.6\linewidth]{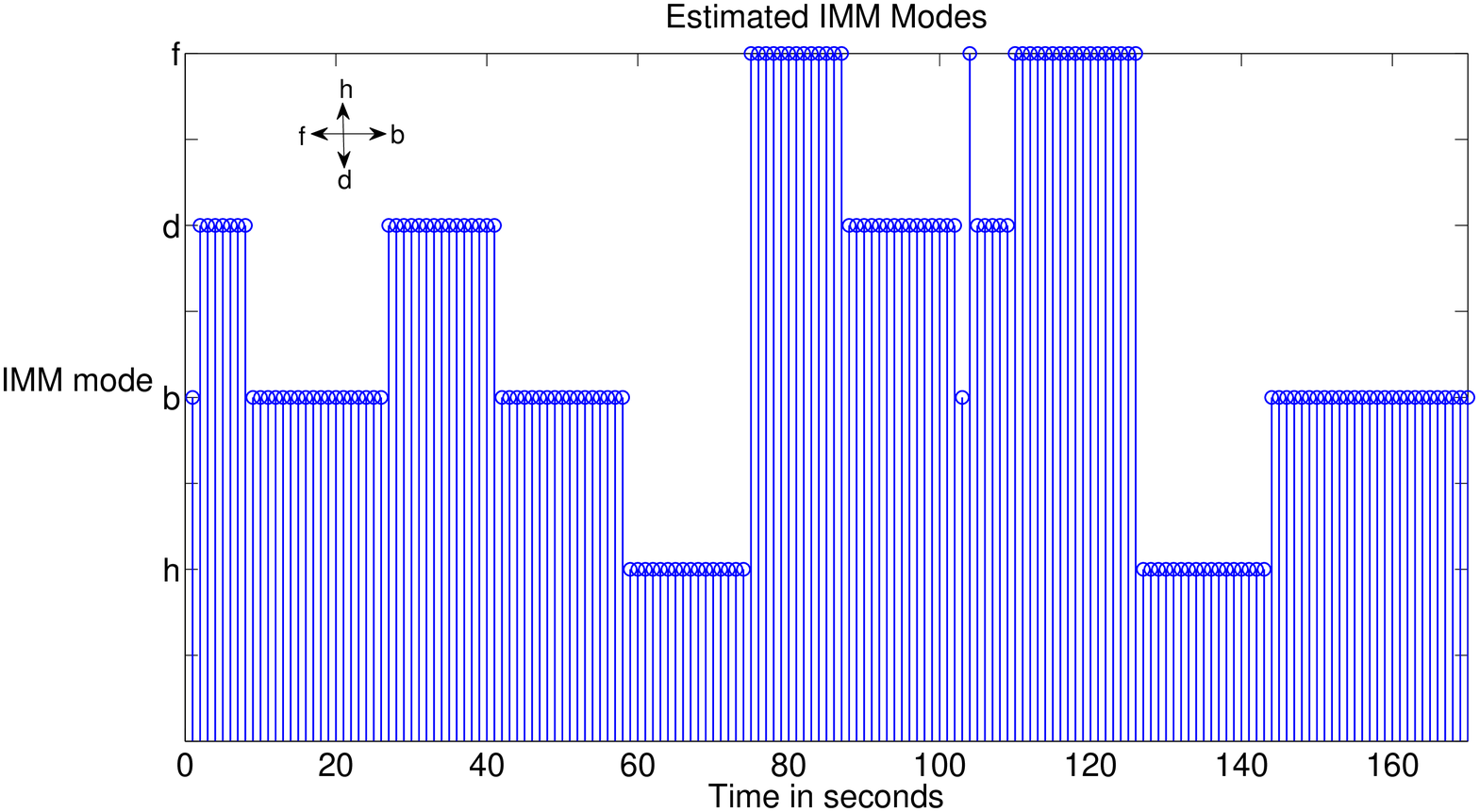}
%\end{minipage}
\caption[The output of the IMM/Extended Kalman filter for GMTI tracking]{The output of the IMM/Extended Kalman filter. The result of the particle filter is not shown because it's very similar. The top panel illustrates the real trajectory of the truck, and the track developed by the filter. The bottom panel, on the other hand, shows the estimated IMM modes. The set of IMM modes corresponds to the set of terminals that is to be parsed by the algorithm for the identification of the geometric pattern. }
\label{GMTI:fig:estimatedIMMFull}
\end{figure}

The performance of the syntactic filtering is illustrated by dealing with two geometric patterns: an arc pattern in a pincer scenario, and a m-rectangle in loitering situation. Numerical studies are done with both the particle filter and the IMM/extended Kalman filter, but since the results are very similar, only the results of the IMM/Extended Kalman filter is shown. The tracking result illustrated in Fig. \ref{GMTI:fig:estimatedIMMFull} is based on  a run of the DRDC flight trials. The solid line of the figure on the top is the real GMTI track, and the dotted line is the output of the IMM/Extended Kalman filter. It can be observed that the tracker performs quite well even during the turns of the truck trajectory. An intuitive explanation for this performance is the constraints imposed by the IMM modes $a_k$. Since the mode constrains the noise term and thus reduces the uncertainty of the state estimates, a better estimate of the track is expected even at the turns. 

The IMM/Extended Kalman filter generates the terminals for syntactic parsing, which, as described in Sec. \ref{subsec:GMTIprimitive}, corresponds to the IMM modes. The bottom panel in Fig. \ref{GMTI:fig:estimatedIMMFull} shows the estimated IMM modes, and only four modes are shown for easy display. The syntactic parsing of the IMM modes could be either soft or hard (as in soft or hard decision making). Hard parsing parses the estimated IMM modes, and soft parsing parses the probabilities of the IMM modes. We focus mainly on soft parsing, and numerical results of parsing the arc and the square pattern are shown next.

\begin{figure}
\centering
\includegraphics[width=0.8\linewidth]{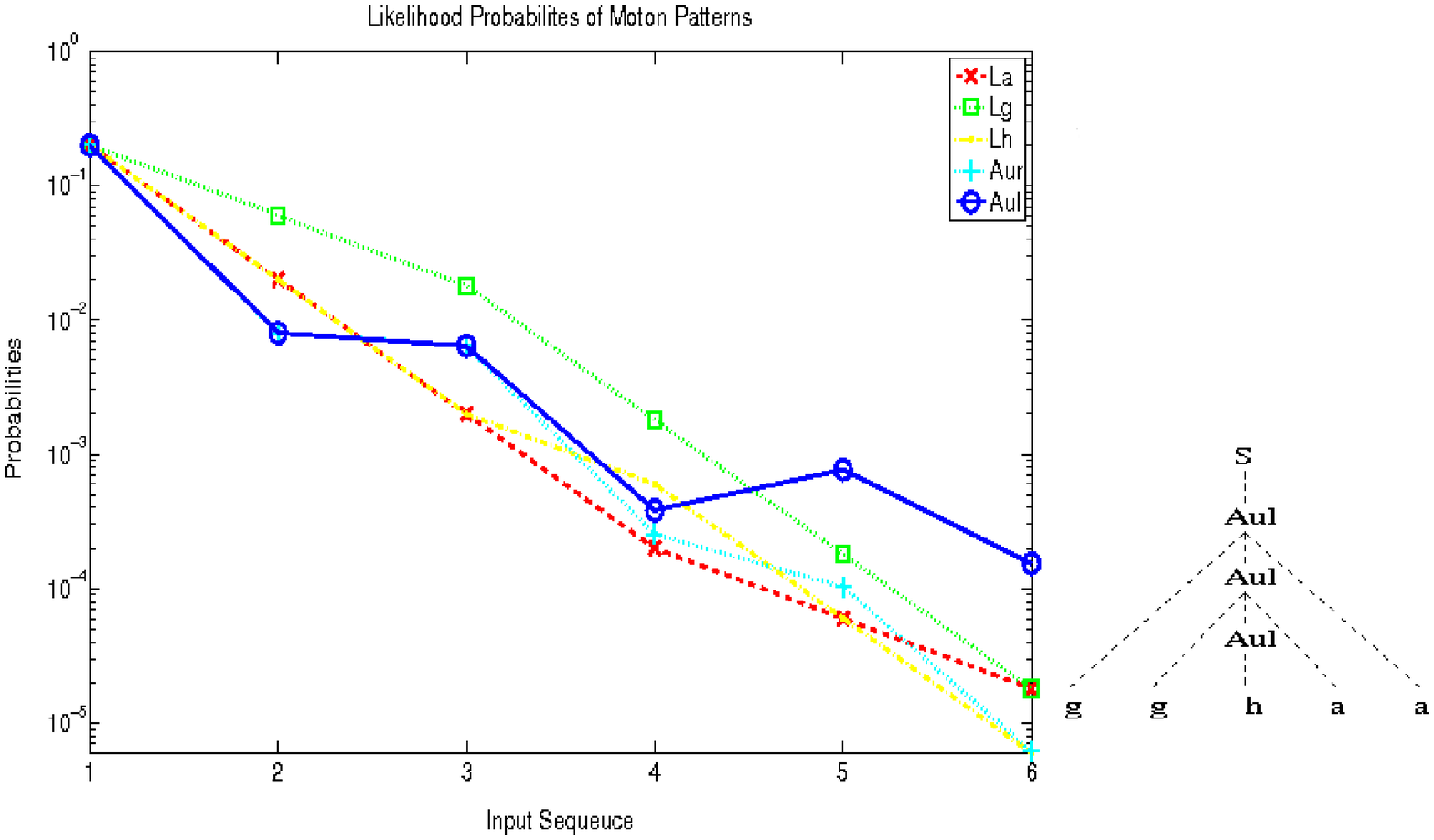}
\caption[Likelihood probabilities of geometric patterns from ground target tracking]{The plot demonstrates the likelihood probabilities of different geometric patterns as the input sequence of IMM modes corresponding to an arc is being parsed.}
\label{GMTI:fig:simresult}
\end{figure}

\begin{figure}
\centering
\includegraphics[width=0.6\linewidth]{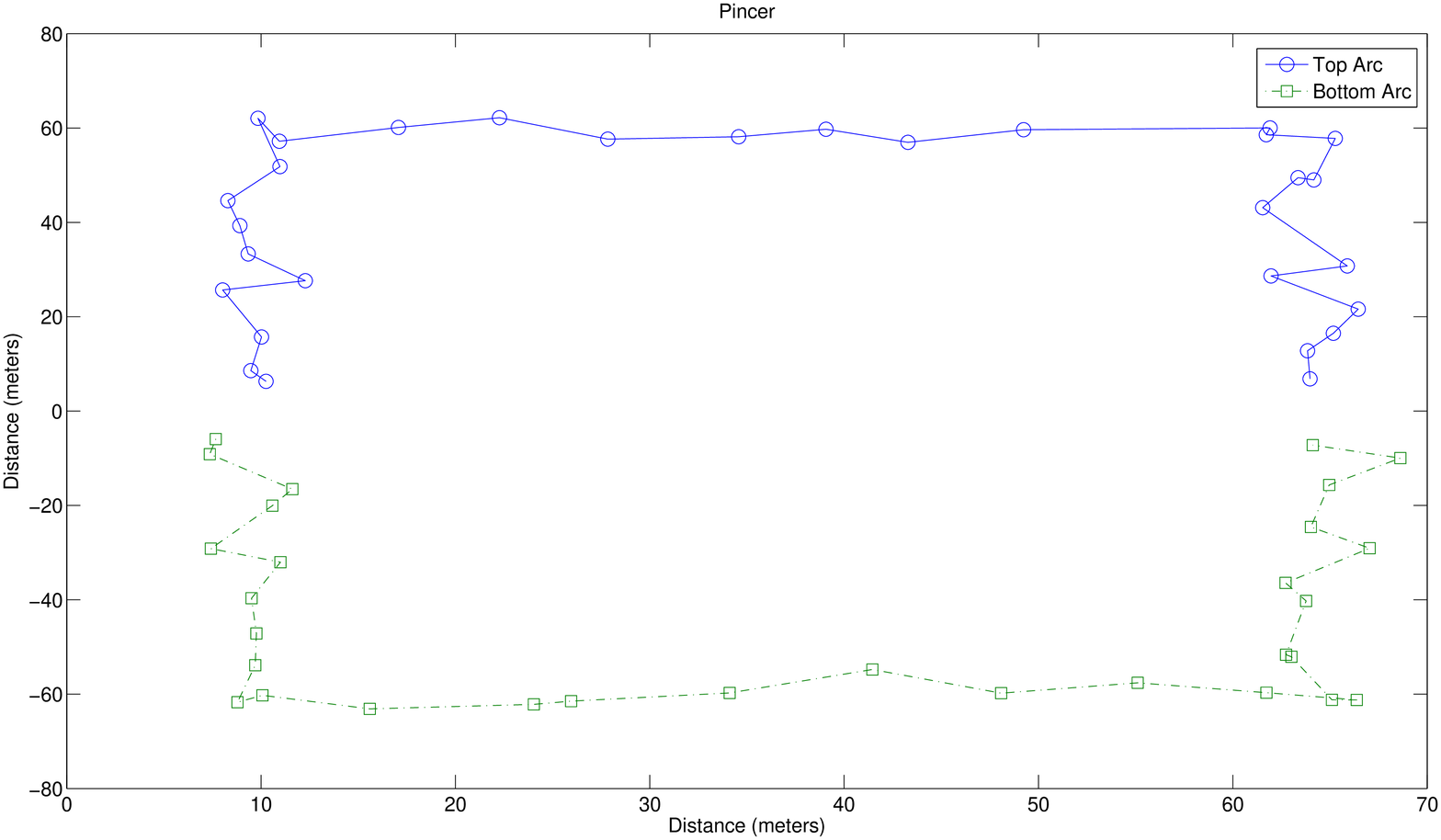}
\caption{The trajectories of a pincer operation.}
\label{GMTI:fig:pincer_trajectory}
\end{figure}

Fig. \ref{GMTI:fig:simresult} shows the likelihood probabilities of different geometric patterns as an arc is parsed, and the most likely parse tree. The parsing algorithm initially classifies the trajectory as a line, but as more data arrives, it correctly identifies the trajectory as an arc. Fig. \ref{GMTI:fig:pincer_trajectory} shows two arcs in the pincer trajectory. The detection data arrived not as two independent tracks, but an an out of order interleaved sequence. The parsing algorithm performs the data association as described in Sec. \ref{subsec:earleyparser}, and parses the two arcs separately. %Once the arcs are recognized, the identification of the pincer can be easily achieved by some logic operator. 
It should note that an arc is a palindrome and it is important to identify an arc irrespective of its dimension and orientation. 

\begin{figure}
\centering
%\begin{minipage}{0.6\linewidth}
\includegraphics[width=0.8\linewidth]{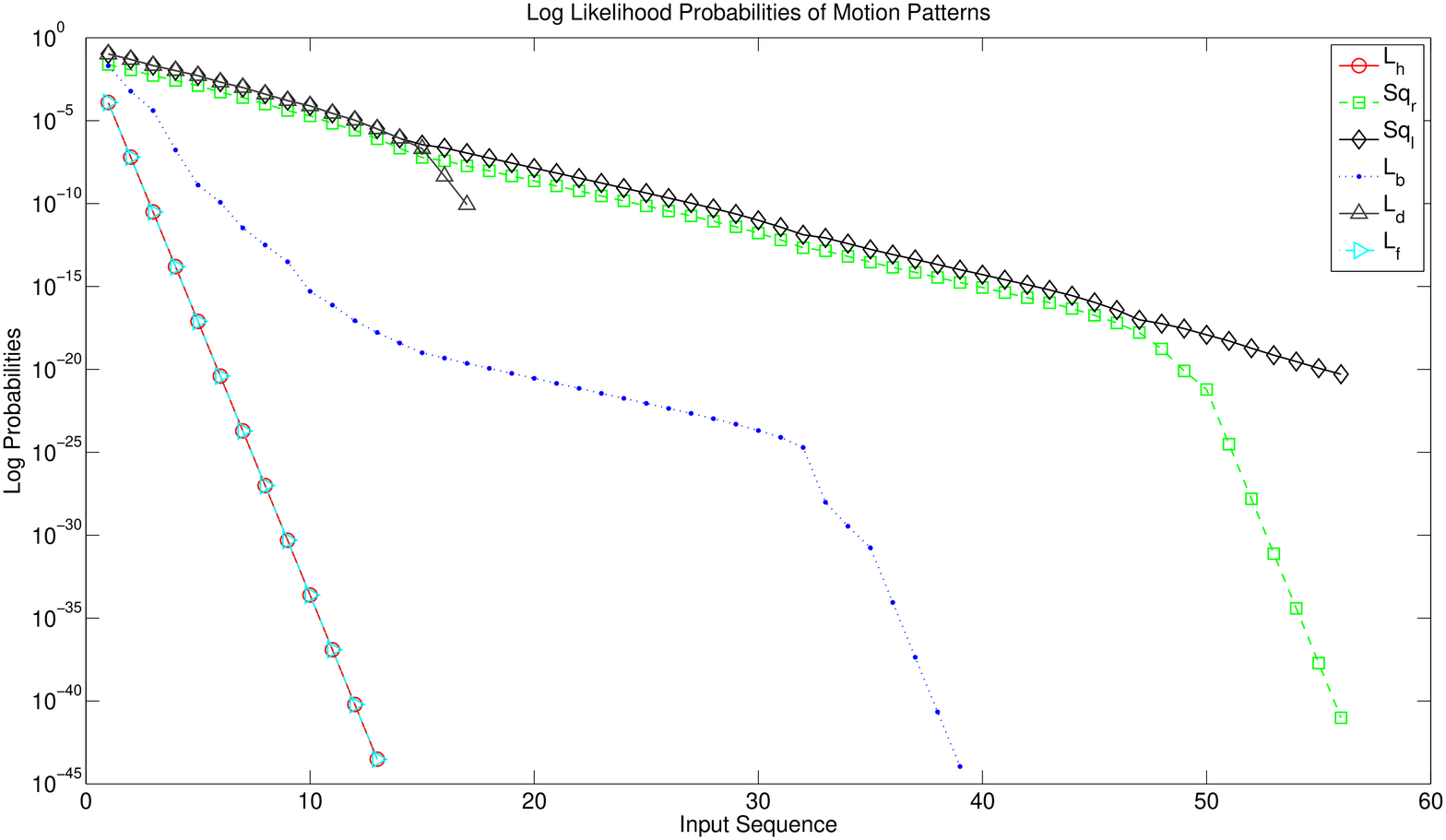} \\
%\end{minipage}
\begin{minipage}{\linewidth}
\begin{minipage}{0.48\linewidth}
\includegraphics[width=\linewidth]{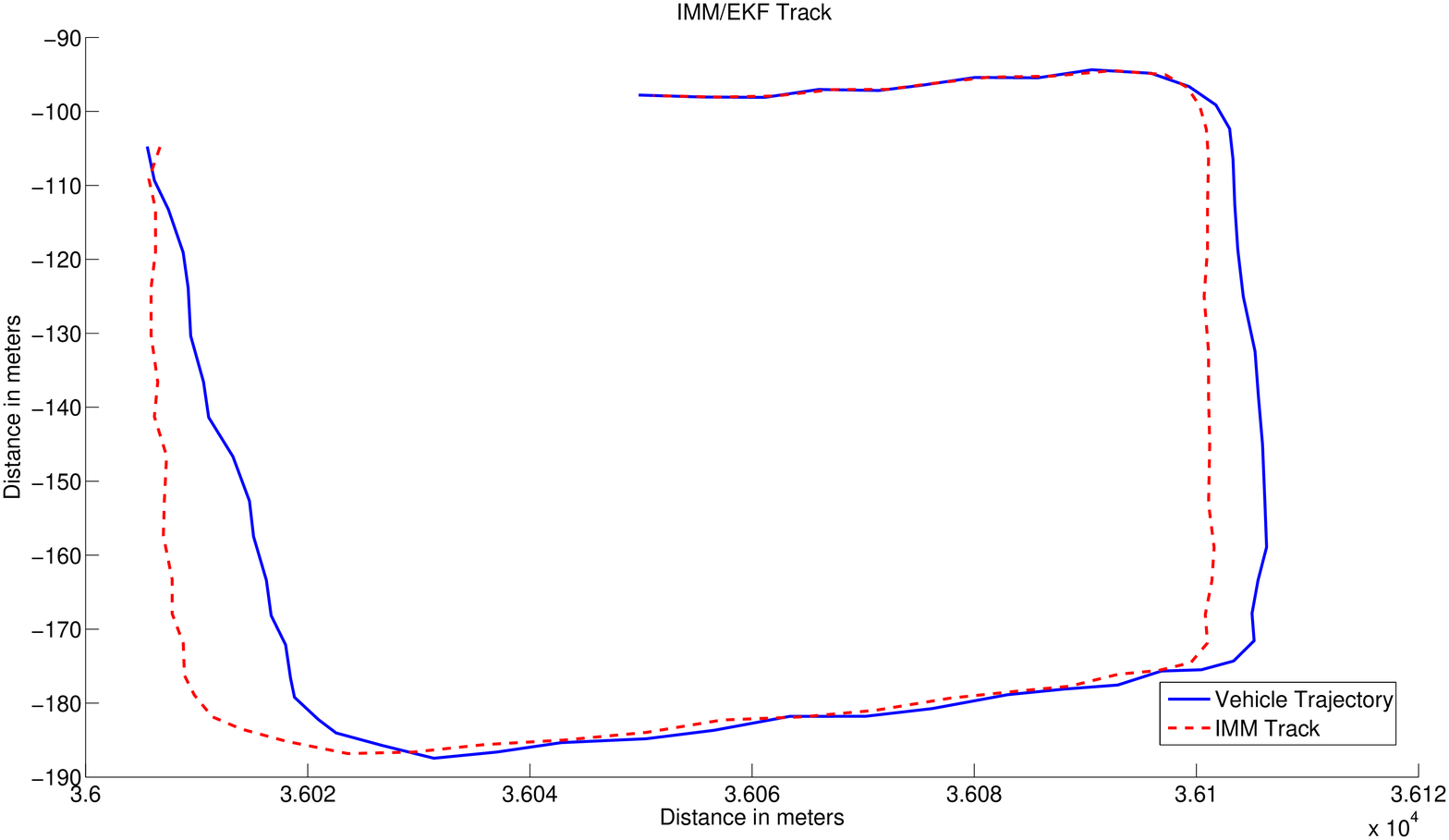} \\
\end{minipage}
\begin{minipage}{0.48\linewidth}
\includegraphics[width=\linewidth]{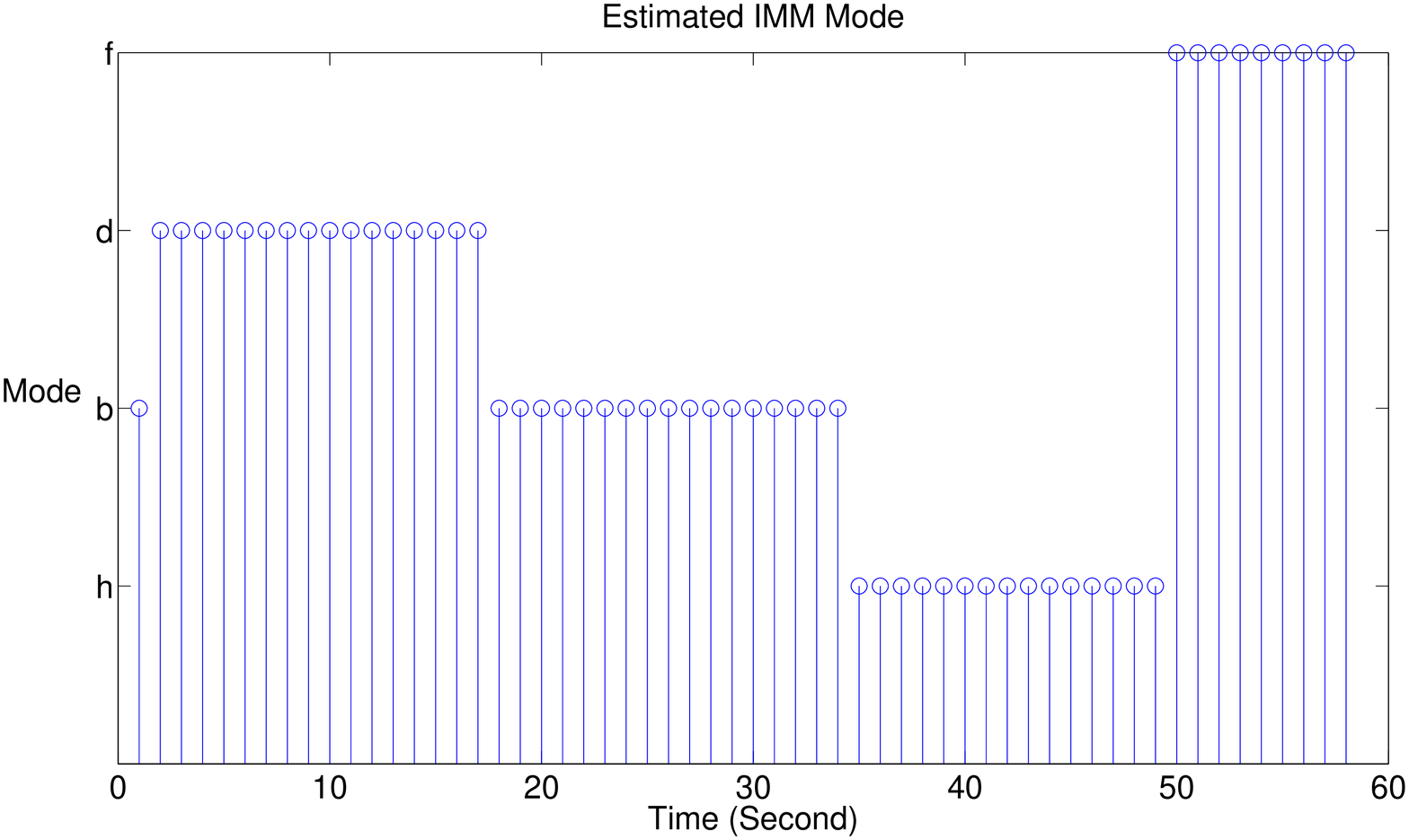}
\end{minipage}
\end{minipage}
\caption[Numerical results of parsing a square pattern]{The figure illustrates the numerical result of parsing a m-rectangle pattern. The log likelihood probabilities of different geometric patterns are shown in the top figure. The trajectory and its corresponding track are shown at the bottom left figure, and the estimated IMM modes are shown at the bottom right figure. }
\label{GMTI:fig:parsedIMM}
\end{figure}

Fig. \ref{GMTI:fig:parsedIMM} illustrates the likelihood probabilities of different geometric patterns as an m-rectangle is parsed. We used a much longer track in this study to demonstrate the practicality of the algorithm. However, the parse tree is omitted due to its large size. As it can be seen from the top panel of the figure, the correct geometric pattern maintains its high probability as the probabilities of other patterns drop because the input sequence does not support them. Some patterns such as vertical line and clockwise m-rectangle had high probabilities initially because the initial segment of the input terminal string matches their syntactic structure. However, as more terminals are parsed, their probabilities drop. This observation means that it is possible to prune a parse tree as its probability drops below a certain threshold. If the input terminal sequence does not support the syntactic rules of a syntactic pattern, the parse tree corresponding to the pattern could be pruned completely, and which could greatly reduce the computational complexity and the storage requirement.

\subsection{Performance of Syntactic Enhanced Tracker}
\label{GMTI:subsec:feedback}

Above parsing results demonstrate how SCFG signal processing can estimate the geometric patterns of the target trajectories. A natural question is: {\em Can the syntactic tracker estimates be fed back to the standard tracking algorithm to improve performance}? For example if the syntactic tracker estimates that the target is moving in an arc, this information should be useful to the lower level tracking algorithm.

\begin{figure}
\begin{minipage}{0.3\linewidth}
\includegraphics[width=\linewidth]{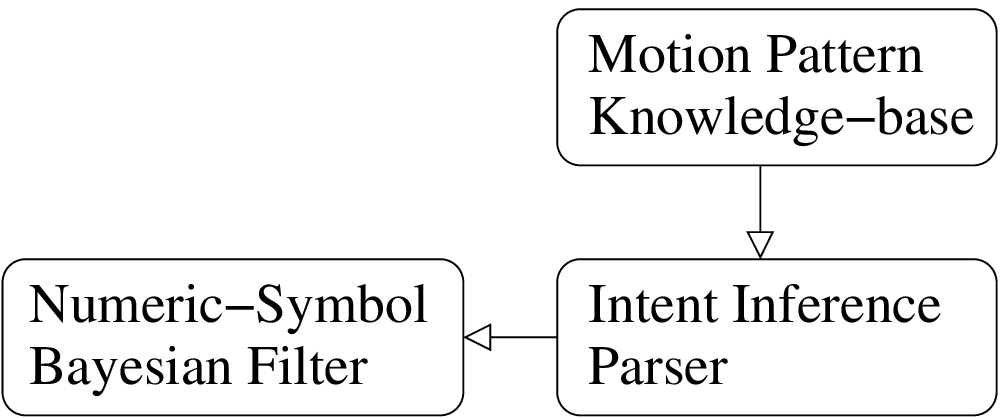}
\end{minipage}
\begin{minipage}{0.55\linewidth}
\includegraphics[width=\linewidth]{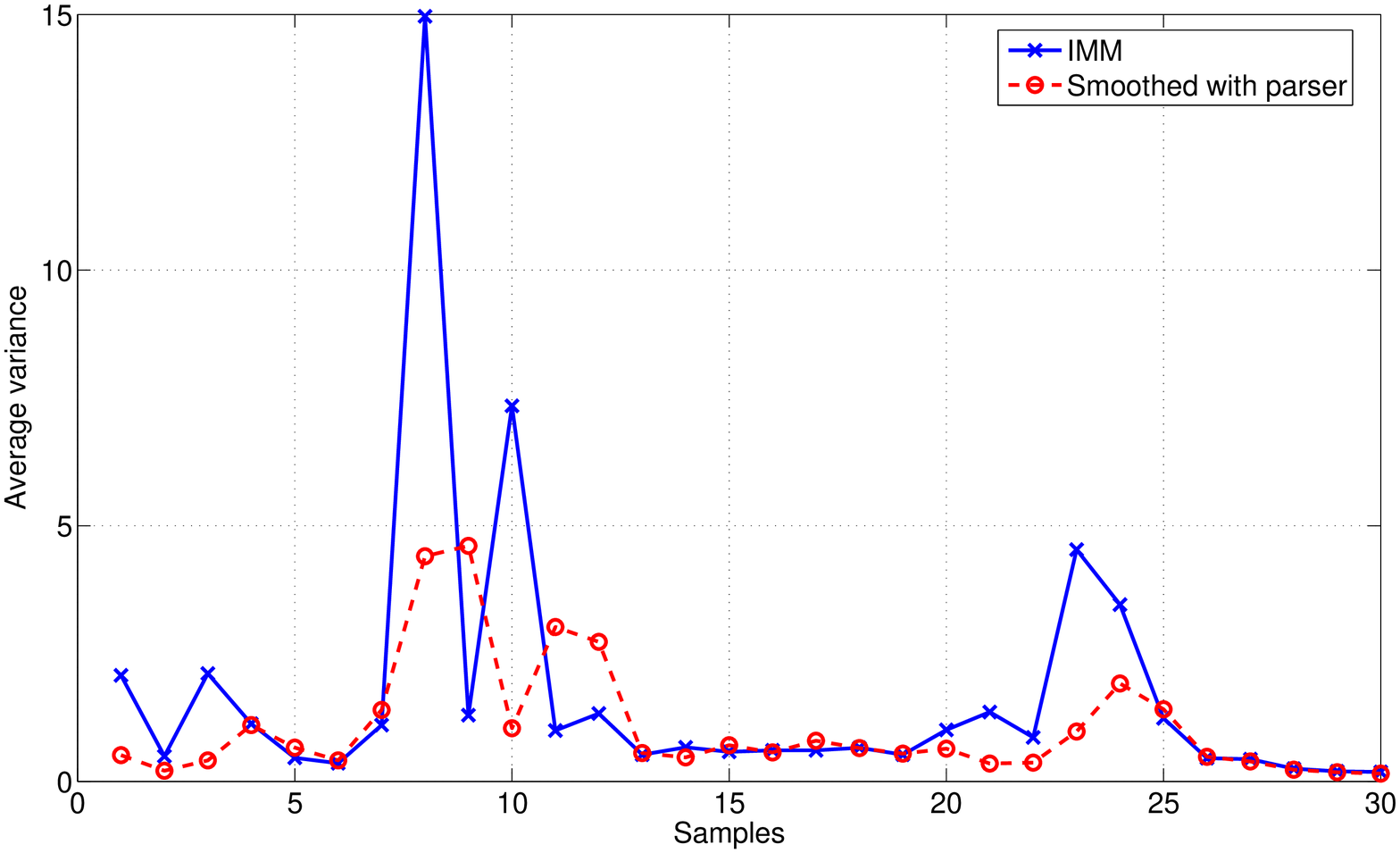}
\end{minipage}
\caption[Tracking covariance reduction from feeding back syntactic description]{The figure shows the covariance reduction from feeding back the syntactic level description to the Bayesian tracking module.}
\label{GMTI:fig:covariance}
\end{figure}

We used the syntactic tracker of Sec. \ref{subsec:earleyparser} and fed the estimates to the multiple mode Bayesian filter using (\ref{eqn:feedback}), where the mode probability is computed as the weighted sum of the IMM mode estimates and the SCFG parser estimates. The SCFG parser calculates the probability $P(a_k|a_{1:k-1}^*,G^{CFG},z_{1:k})$ based on the outputs of the prediction states of Earley-Stolcke parser at each time instant (Detail of the computation can be found in \cite{Jel97}). Since the IMM and the SCFG offers complimentary information of the mode, we mix the two models equally for each mode estimate, i.e., $P(G^{CFG}|z_{1:k}) = P(G^{RG}|z_{1:k})=0.5$. Fig. \ref{GMTI:fig:covariance} demonstrates the reduction in estimator covariance with knowledge of the extracted geometric pattern. The solid line shows covariance of the tracker as the target is moving in a m-rectangle, and the dotted line shows covariance of the assisted tracker. The jumps in covariance correspond to the times when the target is making sharp turns, and knowledge about the target trajectory's geometric pattern allows the tracker to make better predictions of the turns, and thus reduce covariance. 

\section{Conclusion}
\label{sec:conclusion}

In this paper we considered syntactic (higher-level) tracking  of ground targets using GMTI radar.
The goal of such syntactic filtering is to assist
human radar operators in making inferences about
the target behaviour given track estimates.
Our premise for syntactic signal processing is that the geometric pattern of a target's trajectory
can be modeled as "words" (modes) spoken by a SCFG language.
 The syntactic tracker constructs a  parse tree of the geometric patterns that form
 the target trajectory and provides valuable information about the targets' intent. The parsing of the motion trajectories is implemented with Earley Stolcke parsing algorithm, and we extend its control structure with a particle filter and a IMM/Extended Kalman filter to deal with the GMTI data. The parsing algorithm and the Bayesian filters were implemented, and numerical studies are presented using
real GMTI data collected with DRDC Ottawa's XWEAR radar.

\bibliographystyle{IEEEtran}
  

 \end{document}